# Women's visibility in academic seminars: women ask fewer questions than men

Alecia J. Carter[1,2,*], Alyssa Croft[3], Dieter Lukas[1,4], Gillian M. Sandstrom[5]

[1]Department of Zoology, University of Cambridge, Cambridge, United Kingdom

[2]ISEM, Université de Montpellier, CNRS, IRD, EPHE, Montpellier, France

[3]Department of Psychology, University of Arizona, Tucson, United States of America

[4]Department of Human Behavior, Culture, and Ecology, Max Planck Institute for Evolutionary Anthropology, Leipzig, Germany

[5]Department of Psychology, University of Essex, Essex, United Kingdom

[*]Author for correspondence: Institut des Sciences de l'Évolution, Université de Montpellier, Place Eugène Bataillon Cc 065, Montpellier France 34095, T: +33 (0) 4 67 14 31 33,
E: alecia.carter@umontpellier.fr

**Women's visibility in academic seminars: women ask fewer questions than men**

Running title: Women's visibility in academic seminars

## Abstract

The attrition of women in academic careers is a major concern, particularly in Science, Technology, Engineering, and Mathematics subjects. One factor that can contribute to the attrition is the lack of visible role models for women in academia. At early career stages, the behaviour of the local community may play a formative role in identifying ingroup role models, shaping women's impressions of whether or not they can be successful in academia. One common and formative setting to observe role models is the local departmental academic seminar, talk, or presentation. We thus quantified women's visibility through the question-asking behaviour of academics at seminars using observations and an online survey. From the survey responses of over 600 academics in 20 countries, we found that women reported asking fewer questions after seminars compared to men. This impression was supported by observational data from almost 250 seminars in 10 countries: women audience members asked absolutely and proportionally fewer questions than male audience members. When asked why they did not ask questions when they wanted to, women, more than men, endorsed internal factors (e.g., not working up the nerve). However, our observations suggest that structural factors might also play a role; when a man was the first to ask a question, or there were fewer questions, women asked proportionally fewer questions. Attempts to counteract the latter effect by manipulating the time for questions (in an effort to provoke more questions) in two departments were unsuccessful. We propose alternative recommendations for creating an environment that makes everyone feel more comfortable to ask questions, thus promoting equal visibility for women and members of other less visible groups.

## Introduction

Women account for 59% of undergraduate degrees, but only 47% of PhD graduates, 45% of fixed-term contract postdoctoral researchers, 37% of junior and 21% of senior faculty positions across all academic subjects in Europe [European Commission, 2015 [Fig 6.1] http://ec.europa.eu/research/swafs/pdf/pub_gender_equality/she_figures_2015-final.pdf; see also: 1]. The decreasing representation of women in academia as careers progress is frequently referred to as the "leaky pipeline" [2]. Many factors have been proposed to explain the attrition of women as academic careers progress, including innate differences in ability; differences in the career preferences of men and women; the assessment of women's CVs for hiring, tenure and promotion; differences in males' and females' salaries for equivalent positions; parenting; imposter syndrome; and a lack of appropriate role models and mentors for women, all of which lead to reduced visibility of women in academic science [reviewed in 1,3].

Social role theory provides a framework to understand how various factors might influence individuals' decisions to choose an academic career. According to social role theory [4], people tend to make inferences about which characteristics are needed to be successful in a given role by examining the characteristics of the people who most predominantly occupy that role. Because women are often underrepresented in the later career stages in academic science, it is possible that women (and other underrepresented minority groups) might infer that they do not possess or want to express the relevant characteristics for senior faculty positions and therefore do not belong in those particular careers, as has been shown in the medical field [5]. Furthermore, when people do not have first-hand knowledge of their own level of performance in a given domain, they look to the performance of similar others (i.e., ingroup members; in this case other women) to gauge their own potential likelihood of success in that domain [e.g. 6–8]. For these reasons, observing successful models, with whom one can easily relate, is critical for encouraging larger numbers of underrepresented group members to enter and remain in that field [9]. In the case of the "leaky pipeline" for women in academic science, then, the degree to which other women are visible becomes an important problem that needs to be addressed.

In addition to a general pattern of gender inequality in academic posts, women and men—and their contributions—may not be equally visible or equally valued. For example, men are overrepresented in terms of authorship, especially first, senior, and sole authorship [10–13] and men's papers are cited more often [10,14]. In addition, when considering contributions to individual papers, women were more likely than men to be credited with performing the experiments (i.e., the more physical part of the process), whereas men were more likely than women to be credited with data analysis, experimental design, contributing tools, and writing (i.e., the more conceptual parts of the process) [15]. Just as many factors have been proposed to explain the leaky pipeline, various factors have been cited to explain these differences in the representation of women and men in academia.

For example, the difference in citations has been explained in part by the fact that women cite themselves less often than men do, and men cite other men more than they cite women [14].

Although publications represent one form of conceptual "visibility" for scientists, there are many other forms, including some more literal. Direct interactions involving groups of scientists are likely to have a stronger influence on shaping an individual's impression of the academic community. One forum where this occurs is at international conferences, where differences in visibility are known to occur: women are less likely than men, and less likely than expected given their proportional representation in a field, to give talks at conferences, and more likely to contribute to less prestigious (and less visible) alternatives, such as posters [16–18]. Although some part of this underrepresentation may be due to selection bias, other explanations have been proposed; for example, women are more likely to decline invitations to give a talk [18], and more likely to seek out shorter rather than longer talks [19]. Another way in which women are less visible at conferences is in their question-asking behaviour: a small number of studies have reported that women ask proportionally fewer questions than men at these events [20–22].

In this study, we examine a form of visibility that is more common and frequent, and apparent earlier in the pipeline (i.e., to junior academics): question-asking behaviour at local departmental academic seminars (i.e., talks, presentations, colloquia, etc.). Social role theory suggests that women should benefit from being exposed to successful ingroup role models at all points along the leaky pipeline. Before attending academic conferences and seeing women present their work, and before gaining a familiarity with the authors of papers in a particular research area, undergraduate and postgraduate students are exposed to the role-modelled behaviours of the women and men who work in their department. Given social role theory explanations for how gendered expectations of certain roles develop based on who is seen occupying those roles, we argue that the behaviour of the local community may play a formative role in identifying ingroup role models at an early career stage. Few studies have investigated such local phenomena, but these reveal a potential bias against women. For example, female undergraduate students are less likely to volunteer to answer an instructor's questions in class, and somewhat less likely to pose their own questions [23]. Such differences in behaviour might emerge through reinforcement: during the early years of schooling girls are slightly more likely than boys to raise their hands to ask a question but teachers are less likely to choose them to answer [24].

Our aims were to determine whether women and men differ in their visibility at academic seminars and which factors might underlie any differences. With regards to the first aim, we tested the hypothesis that women would ask fewer questions at departmental seminars, thus limiting their potential visibility to others. We were interested in individuals' actual question-asking in seminars, to quantify directly any disparity that might exist. With regard to the second aim, we were interested in perceptions of question-asking in seminars, to understand the motivations and beliefs that underlie any disparity. Thus, our data collection also took two approaches. First, we ran an online survey that collected data on over 600 academic respondents' self-reported attendance and question-asking in seminars, their perceptions of others' question-asking behaviour in seminars, and their beliefs about why they themselves and others do and do not ask questions in seminars. Second, we collected observational data at almost 250 seminars in 10 countries to quantify the attendance and question-asking behaviour of women and men in departmental seminars.

Using these two data sets, we asked three questions. First, we asked whether there was a gender disparity in the question-asking of audience members in academic seminars (Question 1). Using data collected in the survey, we asked academics whether they perceived a disparity in women's question-asking in seminars (Q1a). We also used our observational data to describe women's and men's actual question-asking behaviour at seminars (Q1b). Second, we aimed to understand why there is a disparity in women's question-asking in academic seminars (Q2). Using the survey data, we asked both women and men why they did not ask questions when they wanted to, and for those that thought there was a gender disparity in question-asking, we asked why they believed there to be a disparity (Q2a). Next, we used our observational data to identify factors associated with the disparity (Q2b). Finally, we aimed to explore ways of addressing the disparity (Q3). Based on preliminary findings from the first year of our observational data collection, we ran an experiment in two departments to manipulate the time given to questions, in an attempt to promote a gender balance in the audience's question-asking (Q3a). We also asked the survey respondents what they thought could be done to ameliorate the gender disparity (Q3b).

## Materials and Methods

*Online survey: Seminar participation and perceptions*

The survey received ethical approval from the Science and Health Faculty Ethics Sub-Committee of the University of Essex. Participants declared their consent prior to participation and could withdraw from the survey at any time or leave any question unanswered. After completion, participants were briefed about the purpose of the study and provided with contact information in case they wanted further details. No identifying information was collected during the survey, and all data were pooled prior to analyses. To ensure data privacy, the survey was administered through Qualtrics (from an institutional account at the University of Cambridge).

The survey was advertised via social media (Twitter, Facebook) and emails to relevant academic groups, and was active between 16th June 2016 and 22nd August 2016.

The survey asked for details on the participants (gender, academic subject, career stage, country), the structure of academic seminars at their institution (e.g., typical length of time for questions), and their own attendance and question-asking behaviour at seminars. Finally, we asked for their impression of any gender disparity in question-asking and potential reasons for it (for the full survey design see Supplementary Material 1). We disguised our specific interest in a gender disparity by also asking whether question-asking behaviour was related to seniority, confidence, extraversion, and competence. Data on these distractor questions were not analysed in this study.

*Observation of seminar participation*

To determine the extent of the gender disparity in question-asking during academic seminars, we observed seminars and recruited colleagues through personal contact to do the same. Because these data were collected passively at public events, ethical approval was not needed (following https://memforms.apa.org/apa/cli/interest/ethics1.cfm). Observers were in the same fields as the authors (biology or psychology), chosen to represent as much geographic distribution as possible; they were based in 10 different countries and 35 different institutions. We solicited observers' help by explaining the motivation for the study and our preliminary findings (see Supplementary Material 2). In the end, more than 90% of people that were invited to act as observers reported observations. Data were collected opportunistically during seminars that the observers normally attended in their institutions and these seminars are therefore likely to be a representative sample of the broader experiences of academics.

We provided all observers with written guidelines prior to the start of their observations (see Supplementary Material 2). During the initial period of observations at the University of Cambridge, two of us (AJC and DL) attended six seminars together but independently scored them. This yielded identical observations regarding the gender of the first person to ask a question and the total number of questions asked by each gender, and the counts of the audience numbers were within 0-2 people, suggesting that the guidelines are sufficiently specific for comparison across observers.

For each seminar, observers recorded: whether the speaker was an external visitor or affiliated with the hosting institution; the gender of the speaker; the start and end time of the presentation, and the start and end time of the question period after the presentation; the number of women and men in the audience; the number of questions asked by women and by men; and the gender of the person asking the first question. Each observer recorded the number of women and men among the faculty of the hosting department based on the teaching staff listed on the institution's official website.

We recorded gender as perceived by the observer. This is likely to reflect the perception of other audience members, but we acknowledge that this may not match the target's gender identity. As we wanted a measure of the potential opportunities for the visibility of each gender, observers recorded the total number of questions (including multiple questions from the same person), rather than the total number of different people asking questions. This is because after most talks, there is a limited amount of time for questions; multiple questions asked by the same questioner therefore raises the visibility of that particular gender in proportion to the number of questions asked.

*Experimental manipulation of time given to questions*

We (AJC and DL) collected preliminary observations of question-asking from the University of Cambridge during the 2014-2016 academic years ($N$ = 62, comprising 18, 18, and 26 seminars in each of three departments). These data indicated a correlation between the number of questions asked and the imbalance in questions, with the imbalance approaching 0 as more questions were asked (linear mixed effects model with department as a random effect: $\beta \pm$ S.E. = 0.02 ± 0.009, $t$ = 2.02). Based on this preliminary finding, we hypothesised that we could increase the number of questions asked by women by increasing the amount of time devoted to questions after seminars. We thus designed a manipulation at two institutions to test whether decreasing the length of talks (and thus, theoretically, increasing the time allotted to questions) would lead to more equal question-asking from male and female audience members. While these seminar series previously had indicated to speakers that presentations should last for about 45 minutes, during the manipulation we asked speakers in the invitation email to plan for their talk "to last for 40 min with 20 min for questions. This format is designed to encourage a more discursive and inclusive question session in our department."

*Data and analyses*

Our data and analysis scripts are available in the institutional repository of the Max Planck Society at https://dx.doi.org/10.17617/3.12. Our analyses were conducted in R v3.2.2. For each, we list the approach and specifications in the results below. Generalised and linear mixed models were analysed using the lme4 package [25]; because this package does not report $p$-values for linear mixed models, we considered $t$-values over 1.94 as statistically significant and report these below.

## Results

*Descriptives*

In total, 600 people provided consent and started our online survey, and 518 (90%) recorded a response when asked their gender (the last question in the survey), including 303 (58%) women, 206 (40%) men, 4 transgender/non-binary, and 5 who preferred not to report their gender. We restricted our analyses to the responses of women and men given the small number of respondents who did not consider themselves within these categories, resulting in a sample of 509 responses for our analyses. Survey respondents were from the academic community: 2% were undergraduates ($N$ = 12), 38% were post-graduates ($N$ = 192), 20% were postdoctoral researchers ($N$ = 102), 5% were research fellows ($N$ = 26), 29% were faculty ($N$ = 150), 5% were "other" ($N$ = 27). The participants who completed the online survey were from 19 different countries (9 participants did not provide information about country) and 28 fields of study (28 participants did not provide information about field of study). The majority of respondents who indicated their field of study (74%; $N$ = 356) were from the same fields as the authors of this study: biology and psychology.

Observational data were collected at 247 seminars, from 42 departments of 35 institutions in 10 countries. We retained the pilot data collected at the University of Cambridge and the seminars that were subject to the experimental manipulation, since we found no effect of our manipulation on the time given to questions (see below).

*The current culture of academic seminars*

We first aimed to describe the general patterns of academics' attendance at and question-asking in departmental seminars. Overall, most people reported in the online survey that they attended seminars weekly ($N$ = 200, 35%) or fortnightly/bi-weekly ($N$ = 143, 25%). On average, there were 34 people in attendance at the seminars that were observed (range = 6-220, IQR = 25-46). The majority of seminars ($N$ = 113; 47%) started between 16:00 and 16:59, and attendance increased slightly for seminars starting later in the day compared to earlier in the day (generalised linear mixed model with department nested within the university as a random effect and a Poisson error structure for count data: β ± S.E. = 0.061 ± 0.015, $z$ = 4.19, $p$ < 0.001). On average, we observed 6 questions (range = 0-24, IQR = 4.5-8) over 12 min of question time (range = 2-60, IQR = 10-17.5).

*Gender differences in attendance and question-asking behaviour*

There was no difference between men and women in self-reported frequency of attendance, $\chi^2(4)$ = 1.58, $p$ = 0.82, or observed attendance (average proportion of women attendees = 0.51, range = 0.14-0.78, IQR = 0.43-0.59; $t$-test of whether the proportion is different from 0.5: $t(245)$ = 1.54, 95% CI = 0.50, 0.53, $p$ = 0.13). However, fewer women than men were seminar speakers ($N_{female}$ = 100, $N_{male}$ = 147; exact binomial test of whether the probability of a female speaker is different to 0.5: observed proportion = 0.40, 95% CI = 0.34, 0.47, $p$ = 0.003). Seminars later in the day were attended by proportionally more women than those earlier in the day (linear mixed effects model with the proportion of women as the response, and hour of the day as the predictor, with department nested within the university as a random effect: β ± S.E. = 0.017 ± 0.0055, $t$ = 3.01).

In general, men and women did not differ in their motivations for asking questions; approximately equal proportions of men and women reported being motivated by an interest in the subject (92% of men; 92% of women), the need for clarification (67% of men; 64% of women), the desire to act as a model for more junior academics (32% of men; 31% of women), or to establish a connection with the speaker (26% of men; 30% of women), $t$'s < 1.10, $p$'s > 0.25. However, approximately twice as many men (33%) as women (16%) reported being motivated to ask a question because they felt like they spotted a mistake, $t(362.2)$ = 4.61, $p$ < 0.001 (note: degrees of freedom adjusted due to violation of the assumption of homogeneity of variances, as indicated by a significant Levene's test).

Most respondents reported that meeting a speaker informally was actively encouraged in their department ($N$ = 219; 42%) or required only contacting the host ($N$ = 185; 36%). However, men and women differed in their perceptions of the availability of a speaker, $\chi^2(3) = 14.50$, $p = 0.002$, with female respondents reporting twice as frequently as men that the speaker only met with relevant faculty (24% of female vs 11% of male respondents).

*Q1: Is there a gender disparity in participation in academic seminars?*

We aimed to quantify whether academics perceive a gender disparity in the proportions of men and women who ask questions in seminars, and whether this perception differs according to gender. Most respondents reported that gender played a role in who asked questions at seminars, reporting that they believed that men were more likely to ask questions (*N* = 279 (58%); see Fig 1C). However, men and women differed in their endorsement of this belief; women reported more frequently than men that they believed there was a bias towards men asking questions (*N* = 182 women (60%) vs. 97 men (47%); $\chi^2(2) = 8.40$, $p = 0.01$).

These perceptions about a gender disparity in question-asking were borne out by the self-report data. Men and women differed in how frequently they reported asking questions, $\chi^2(4) = 21.71$, $p < 0.001$: women self-reported asking questions less frequently than men (see Supplementary Material 4 results, Fig S4.1B). Despite this, the vast majority of respondents of both genders reported that they sometimes did not ask a question when they had one (*N* = 277 women (91%); 189 men (92%); overall 92%).

We next examined whether the observational data substantiated these perceptions and self-reports of a disparity in women's question-asking after seminars. To test whether the proportion of questions asked by women differed from the proportion of women present in the audience, we ran a two-tailed t-test comparing the difference in these proportions to 0 (no difference). Survey respondents' (especially women's) general belief that men ask more questions than women was supported by the actual behaviour observed in seminars: proportionally fewer women asked questions after seminars than would be expected given the proportion of women in the audience ($M = -0.19$, 95% CI = -0.16, -0.22, $t(245) = -12.55$, $p < 0.001$, Fig 1A,B). Put another way, male attendees were over two and half times more likely to ask a question than women attendees (odds ratio = 2.57) during the seminars that we observed.

**Fig 1 title:**

The gender "disparity" in question-asking

**Fig 1 caption**: A (a) scatter plot of the proportion of women in the audience plotted against the proportion of questions asked by women after a seminar, (b) a histogram of the size of the disparity at each seminar, and (c) a barchart of the beliefs of each gender about whether there is a disparity. Panel (a) shows a visual representation of the disparity in the proportion of questions asked by women (i.e., the difference in the proportion of women in the audience and the proportion of questions asked by women). Points falling in the lower orange half of the plot indicate a disparity towards men, whilst points falling in the upper green half indicate a disparity towards women audience members. Indicated are two seminars that fall in different categories. The green arrow indicates a seminar with a bias towards questions from women, in which the proportion of women in the audience was 0.38, but the proportion of questions asked by women was 0.67. Conversely, the orange arrow indicates a seminar with a bias towards questions from men, in which the proportion of women in the audience was 0.78 but the proportion of questions asked by women was 0.40. Panel (b) shows the frequency at which the disparities were observed, with orange bins indicating seminars with questions disproportionately asked by male audience members and green bins indicating seminars with questions disproportionately asked by female audience members. In both panels, the red line indicates no disparity (i.e., the proportion of the women in the audience matched the proportion of questions asked by women). Panel (c) shows the proportions of female (green) and male (orange) respondents who indicated that they believed that men or women asked more questions in seminars, or that questions were asked equally by men and women.

*Q2: Why is there a disparity in question-asking behaviour?*

We next aimed to understand why there is a disparity in women's question-asking at seminars. The vast majority of our online survey respondents (91% of women and 92% of men) reported sometimes not asking a question when they had one. We asked them what prevented them from asking a question in these cases on a Likert scale from 1 (not at all important) to 5 (extremely important). The results are summarised by gender in Table 1 (for detailed results, see Table 1 in Supplementary Material 3). Overall, men and women differed in their ratings of the importance of each reason for not having asked a question (Fig 2, dark circles; Table 1), except for the reason that they were meeting with the speaker after the seminar (not shown). For example, *not feeling clever enough to ask a question* was rated as more important by women ($M$ = 2.93, $SD$ = 1.38), than by men ($M$ = 2.34, $SD$ = 1.36) (Table 1; Fig 2). Women rated all the reasons as more important than men did (except for a lack of time, which men judged as more important than women) suggesting that women rated 'internal' factors as more limiting than men.

We coded the 106 open-ended responses to this question. Common responses included: worries and personal characteristics (e.g., being soft-spoken, unassertive, feeling unimportant, feeling uncomfortable with the language; $N_{male}$ = 9, $N_{female}$ = 17); consideration for colleagues ($N_{male}$ = 11, $N_{female}$ = 8), the speaker ($N_{male}$ = 7, $N_{female}$ = 3) and the audience ($N_{male}$ = 3, $N_{female}$ = 3); not being picked by the moderator ($N_{male}$ = 1, $N_{female}$ = 11); someone else asking the question ($N_{male}$ = 7, $N_{female}$ = 2); not enough time (for all the questions, or to formulate own question; $N_{male}$ = 0, $N_{female}$ = 8); and fear of judgment from members of the audience (i.e., peers; $N_{male}$ = 2, $N_{female}$ = 5).

**Table 1:**
Responses of a sample of academics who identify as male and female about (1) what factors prevent them from asking a question after a seminar, even when they had a question to ask and (2) what they believed prevented women from asking a question if they had one.

| ***Question*, factor** | ***t*** | **df** | ***p*** | **Female *M* ± SD** | **Male *M* ± SD** |
|---|---|---|---|---|---|
| *How important is each of these factors in stopping you from asking a question?* | | | | | |
| Couldn't work up the nerve | -4.13 | 396.38 | <0.001 | 3.18 ± 1.34 | 2.65 ± 1.36 |
| The speaker was too eminent/intimidating | -4.24 | 433.33 | <0.001 | 2.24 ± 1.10 | 1.83 ± 0.97 |
| Not my field | -1.98 | 378.65 | 0.049 | 2.75 ± 1.16 | 2.53 ± 1.24 |
| Worried that I was not clever enough to ask a good question | -4.63 | 405.13 | <0.001 | 2.93 ± 1.38 | 2.34 ± 1.36 |
| Worried that I had misunderstood the content | -4.23 | 402.99 | <0.001 | 3.41 ± 1.17 | 2.95 ± 1.16 |
| Not sure whether the question was appropriate | -3.78 | 399.79 | <0.001 | 3.20 ± 1.11 | 2.80 ± 1.11 |
| I was meeting the speaker later / asked after the talk had ended | 1.38 | 397.94 | 0.167 | 2.50 ± 1.22 | 2.66 ± 1.23 |
| Not enough time | 2.02 | 381.43 | 0.044 | 2.80 ± 1.13 | 3.03 ± 1.23 |
| *How important do you think each of these factors is in preventing [women] from asking more questions?* | | | | | |
| Can't work up the nerve | -3.44 | 161.55 | 0.001 | 3.43 ± 1.00 | 2.96 ± 1.03 |
| Feel intimidated by the speaker | -1.47 | 160.94 | 0.143 | 3.01 ± 1.05 | 2.81 ± 1.04 |
| Feel they are not an expert | -5.91 | 142.34 | <0.001 | 3.63 ± 0.97 | 2.78 ± 1.16 |
| Believe that they are not clever enough to ask a good question | -4.17 | 157.81 | <0.001 | 3.21 ± 1.13 | 2.57 ± 1.17 |
| Worry that they misunderstand the content | -7.39 | 164.36 | <0.001 | 3.13 ± 1.05 | 2.11 ± 1.04 |
| Are unsure that their questions are appropriate | -4.47 | 146.01 | <0.001 | 3.28 ± 1.01 | 2.62 ± 1.16 |
| Ask questions after the seminar is over | -4.96 | 160.08 | <0.001 | 3.02 ± 1.13 | 2.29 ± 1.08 |

Presented are the factors; the results of a Welch two-sample *t*-test, including the *t*-value (t), degrees of freedom (df) and the *p*-value (*p*); and the means and standard deviations (*M* ± SD) of the responses of respondents who identify as women and men.

**Fig 2 title:**
Mean importance assigned by women and men to (1) each reason why they *themselves* have not asked a question in a seminar when they wanted to, and to (2) each reason men and women believe *women* do not ask questions when they want to

**Fig 2 caption**: Shown are the mean values for women (green) and men (orange) rating how important each factor is in restricting why they themselves did not ask questions when they wanted to (circles). For the respondents who reported a belief that women ask fewer questions than men, shown are the mean values for women (green) and men (orange) rating how important each factor is in restricting women from asking questions when they wanted to (triangles).

We also asked respondents who had indicated a belief that women ask fewer questions than men why they believe that women do not ask more questions. Women rated each reason we asked them about as more important than men did, except being intimidated by the speaker (Table 1; Fig 2; for detailed results, see Table 2 in Supplementary Material 3). For example, *women not feeling clever enough to ask a question* was rated as more important by women ($M$ = 3.21, $SD$ = 1.13), than by men ($M$ = 2.57, $SD$ = 1.17) (Table 1).

Next, using our observational data, we examined potential predictors of a gender disparity in the questions asked after seminars. We used generalised linear mixed effects models with a binomial response, with questions from female audience members coded as cases, and questions from male audience members coded as noncases. To control for repeated measures, all models included the country, and the department nested within the institution as random effects. We did not include the observer as a random effect because most observers collected data within only one department within an institution.

We aimed to test the following fixed effects: the proportion of women in the audience (centred at 0.50), to estimate whether differences in the number of questions asked by women and men reflect differences in individual contributions rather than just their share of the audience; the gender of the speaker (female or male), to understand, for example, whether attendees might feel more comfortable asking a question of a person of the same gender; the gender of the first person to ask a question (female or male) to understand whether a social role model effect might occur within sessions (see below); the total number of questions asked (centred at the median of 6 questions) and the duration of the question time (centred at the median of 12 min) to understand whether perceived or real competition over asking one of the questions limited some individuals; the hour of the day that the seminar started (integer ranging from 10 to 18) as childcare needs differ throughout the day; the proportion of the permanent staff (faculty) in the host department who were female (centred at 0.50) to understand whether gender biases among individuals asking questions were associated with seniority; the number of attendees to understand whether the genders differed in their response to the size of the audience for their question; the field of study (broadly characterised as biology, psychology or philosophy, based on the department in which the talk took place) to understand whether differences in norms or gender roles in different fields influenced participation; and whether the speaker was internal (i.e., from within the department) or not to understand whether familiarity with the speaker influenced who asked a question. Unsurprisingly, there was covariation between the duration of the question time and the number of questions that were asked (generalised linear model with the number of questions as the response and a Poisson link: $\beta \pm$ S.E. = 0.029 ± 0.0022, $z$ = 13.03, $p$ <0.001); we thus used the number of questions rather than the duration for questions, but found qualitatively similar results when using the number of questions asked (see below).

We also included a number of interactions that we predicted *a priori* could contribute to the disparity. Because gender differences in the speakers' behaviour may induce different behaviour from the audience members, we tested whether the speaker's gender also interacted with (a) the total number of questions asked and (b) the number of attendees to affect the gender disparity in the questions asked. In addition, because the first person to ask a question may set the "tone" for the subsequent (disparity in) questions asked, we investigated the interaction between the gender of the first person to ask a question and (a) the total number of questions asked and (b) the gender of the speaker. Such social influence biases have been found in online interactions, where, for example, the tone of the first comment posted influences the tone of subsequent comments [26]. This resulted in a total of four interactions.

Because we had a large number of *a priori* predictors and our modelling approach was exploratory in nature, we used stepwise model simplification to obtain minimal models whose retained components significantly explained the variation in the response (the probability that a question was asked by a female audience member). We thus started with models that included a set of predictors (from those listed above, described below) and interaction terms, and then used backwards elimination of non-significant terms until a minimal model remained that explained the variation in the gender disparity in questions. Then, each dropped term was added back to the final model, one at a time, to check that it remained a non-significant predictor of the gender disparity.

In predicting the proportion of questions asked by women, we could not include the gender of the first person asking a question, since the first person biases the overall gender ratio of questions, in particular when only few questions were asked. We thus ran two sets of analyses using slightly different data and predictors in the starting models to account for this. The first model included the complete dataset and all fixed effects and interactions not including the gender of the first attendee to ask a question. The second model used a reduced dataset, with the first question removed, and included the gender of the first person asking a question as an additional predictor.

Using the complete dataset, we found that the probability that a question was asked by a female audience member was predicted by the proportion of the audience that was female, the proportion of female faculty in the department, the number of questions asked, the gender of the speaker, and whether the speaker was internal or not (Table 2; Figure 3). Using the baseline values of the minimal model (i.e. with the centred values of the continuous variables and the reference category of the categorical variables) the model predicted that the probability that a question was asked by a woman was 29%. The probability increased by 8% from this baseline (to 37%) when the speaker was male, and by 9% (to 38%) when the speaker was internal. The proportion of female faculty in the department had a positive effect on the proportion of questions asked by women in the audience, however this increase was relatively small—a 5% increase in the proportion of female faculty was associated with a 1.5% increase in the proportion of questions asked by women. Similarly, a 5% increase in the proportion of women in the audience resulted in a 1.6% increase in the proportion of questions asked by women. The more questions that were asked resulted in a greater proportion of questions asked by women; compared to the median number of 6 questions, there was a 3.2% increase in the proportion of questions asked by women when 10 questions were asked, and a 7.6% increase when 15 questions were asked. Under generous circumstances i.e. in a department with 50% female faculty, 50% women attendees, a male internal speaker and 10 questions, the minimal model predicts approximately equal numbers of questions from men and women (51% from women).

To deal with the problem of the gender of the first questioner influencing the gender ratio of questions asked, the second model predicted the probability of a question asked by women subsequent to the first question using a modified data set with the first question removed. For example, for a talk that had a male-first question, and totals of 3 questions from women and 5 from males, the new dataset had totals of 3 questions from women and 4 from males, reflecting the numbers of questions asked subsequently to the first one. Additionally, we removed any talks ($N$ = 3) that had only one question—the first one. To control for the factors that affected the proportion of questions asked by women as found in the "complete" model above, our starting model for this second set included the main effects included in the "complete" model of the first set, as well as the main effect of the gender of the first attendee to ask a question and the two interactions involving that effect.

Using the reduced dataset with the first question removed, we found an effect of the gender of the first person to ask a question (Table 2; Figure 3). Using the baseline values of the model (i.e. with the centred values of the continuous variables and the reference category of the categorical variables) the model predicted that the probability that a question subsequent to the first question was asked by a woman was 33%. When the first question was asked by a male audience member, the proportion of subsequent questions asked by women decreased by 6% (to 27%) compared to when the first question was asked by a woman. All other significant effects detected in the first complete model were retained as significant in the reduced minimal model.

As reported earlier, there was significant covariation between the duration of the question time and the number of questions that were asked. When we included in the model the duration of the question time, instead of the total number of questions asked, it was a significant predictor of the proportion of questions asked by women in the reduced dataset ($\beta \pm$ S.E. = 0.013 ± 0.0063, $z$ = 2.041, $p$ = 0.04) but not in the complete data set ($\beta \pm$ S.E. = 0.010 ± 0.0068, $z$ = 1.51, $p$ = 0.13).

**Table 2:**

Predictors correlating with the proportion of questions asked by women

| Data | Predictor | β ± S.E. | *z* | *P* |
|---|---|---|---|---|
| Complete | **Intercept** | **-0.89 ± 0.13** | **-6.86** | **<0.001** |
| | **Proportion of women in audience** | **1.61 ± 0.53** | **3.04** | **0.002** |
| | **Proportion of women teaching staff** | **1.41 ± 0.41** | **3.42** | **0.001** |
| | **Total number of questions** | **0.038 ± 0.013** | **2.92** | **0.004** |
| | **Speaker gender: male**[a] | **0.36 ± 0.12** | **3.11** | **0.002** |
| | **Internal speaker: true**[b] | **0.43 ± 0.14** | **2.98** | **0.003** |
| | Starting hour | -0.006 ± 0.035 | -0.18 | 0.86 |
| | Number of attendees | 0.002 ± 0.003 | 0.84 | 0.40 |
| | Field: philosophy[c] | 0.12 ± 0.21 | 0.56 | 0.574 |
| | Field: psychology[c] | -0.07 ± 0.16 | -0.47 | 0.637 |
| | Male speaker[a] × number of questions | 0.0062 ± 0.025 | 0.24 | 0.81 |
| | Male speaker[a] × number of attendees | -0.0053 ± 0.0049 | -1.099 | 0.27 |
| Reduced | **Intercept** | **-0.72 ± 0.16** | **-4.49** | **<0.001** |
| | ***Gender of the first attendee to ask a question: Male***[a] | ***-0.30 ± 0.13*** | ***-2.41*** | ***0.016*** |
| | **Proportion of women in audience** | **1.61 ± 0.54** | **3.001** | **0.003** |
| | **Proportion of women teaching staff** | **1.29 ± 0.39** | **3.30** | **<0.001** |
| | **Total number of questions** | **0.038 ± 0.014** | **2.68** | **0.007** |
| | **Speaker gender: male**[a] | **0.37 ± 0.13** | **2.96** | **0.003** |
| | **Internal speaker: true**[b] | **0.42 ± 0.15** | **2.82** | **0.005** |
| | Starting hour | 0.013 ± 0.035 | 0.38 | 0.70 |

| | | | |
|---|---|---|---|
| Number of attendees | -0.00086 ± 0.0029 | -0.29 | 0.77 |
| Field: philosophy[c] | 0.17 ± 0.19 | 0.86 | 0.39 |
| Field: psychology[c] | 0.0016 ± 0.16 | 0.010 | 0.99 |
| Male speaker[a] × number of questions | -0.0023 ± 0.027 | -0.088 | 0.93 |
| Male speaker[a] × number of attendees | -0.0083 ± 0.0044 | -1.87 | 0.06 |
| *Male first question[a] × total number of questions* | *0.029 ± 0.026* | *1.12* | *0.26* |
| *Male first question × male speaker[a]* | *0.21 ± 0.25* | *0.84* | *0.40* |

[a] Reference category: female
[b] Reference category: false
[c] Reference category: biology

**Table 2 caption**: Reported are the datasets used (see text for details), the predictor variables, their effect size and the associated standard error (β ± S.E.), the *z*-value, and the *p*-value. The minimal model's predictors are indicated in **bold** type. The terms added in the reduced model are indicated in *italics*. The values for the non-significant terms (i.e., that were dropped during the model simplification procedure), representing the effect size of the terms when they were added back individually to the minimal model, are reported for completeness.

**Fig 3 title**:
The effects predicting the probability of question asked by a female audience member after departmental seminars

**Fig 3 caption**: Plotted are the effects of (a) the proportion of attendees who were women; (b) the proportion of faculty in the department who are women; (c) the total number of questions that were asked; (d) the gender of the speaker; (e) whether the speaker was internal to the department (true) or not (false); and (f) the gender of the first person to ask a question on the proportion of questions asked by women after a departmental seminar. In each panel, the coloured boxes reflect the areas of disparity in the proportion of questions asked by women after a seminar, with the white area representing moderate to no disparity (proportion of questions from women = 0.50 ± 0.10), the orange area representing a disparity towards male audience members asking questions (proportion of questions from women <0.4) and the green area representing a disparity towards female audience members asking questions (proportion of questions from women >0.6). The predicted values are plotted in all cases to control for other effects. In panels (a)-(c), the predicted logistic relationship is plotted in grey with a transparent grey polygon indicating the standard error around the relationship. Likewise, in panels (d)-(f), the predicted values of the effects are plotted as black points with standard error bars.

*Q3: Is there a way to address the gender disparity in question-asking behaviour?*

We asked the survey participants who had indicated that they sometimes do not ask questions how important several factors could be in encouraging them to ask their questions at seminars (Table 3; for detailed results, see Table 3 in Supplementary Material 3). Respondents indicated that the factors most likely to encourage them to ask more questions were having more confidence ($M$ = 3.53) and having an opportunity to ask the question in person ($M$ = 3.48). The factors they thought least likely to encourage them to ask more questions were having a moderator ($M$ = 2.29), or having a better moderator that engages the audience ($M$ = 2.60). Women were more likely than men to think that all of the factors we listed would encourage them to ask more questions (Table 3).

Given our finding that having more questions results in a greater proportion of questions from women, it is of particular interest that neither men nor women who responded to the survey thought that it would be helpful to have a longer time to formulate questions; one-sample t-tests were significantly below the midpoint (of 3: "might help a bit") for both men ($M$ = 2.52, $SD$ = 1.00), $t(186)$ = 5.85, $p < 0.001$ and women ($M$ = 2.74, $SD$ = 0.96), $t(275)$ = 4.19, $p < 0.001$.

**Table 3:**

Responses of a sample of academics who identify as male and female about what factors would encourage them to ask more questions after a seminar

| **Question, factor** | ***M* ± SD** | ***t*** | **df** | ***p*** | **Female *M* ± SD** | **Male *M* ± SD** |
|---|---|---|---|---|---|---|
| *To what extent would each of these factors encourage you to ask more questions?* | | | | | | |
| Confidence | 3.53 ± 1.28 | -5.04 | 162.85 | <0.001 | 3.89 ± 1.16 | 3.08 ± 1.35 |
| A chance to ask in person | 3.48 ± 0.99 | -2.67 | 169.48 | 0.008 | 3.59 ± 0.96 | 3.24 ± 1.01 |
| Seniority | 3.17 ± 1.26 | -6.33 | 160.11 | <0.001 | 3.65 ± 1.17 | 2.65 ± 1.24 |
| A longer time to formulate the question | 2.68 ± 0.98 | -1.67 | 177.42 | 0.096 | 2.74 ± 0.96 | 2.52 ± 1.00 |
| Moderator doing a better job engaging whole audience | 2.60 ± 1.14 | -2.19 | 166.38 | 0.030 | 2.86 ± 1.11 | 2.52 ± 1.17 |
| Nicer speakers | 2.42 ± 1.04 | -1.68 | 176.27 | 0.094 | 2.55 ± 1.02 | 2.32 ± 1.08 |
| More welcoming host | 2.34 ± 1.08 | -2.16 | 179.09 | 0.032 | 2.53 ± 1.09 | 2.23 ± 1.05 |
| Having a moderator to ask the questions | 2.29 ± 1.14 | -4.32 | 184.84 | <0.001 | 2.61 ± 1.10 | 1.98 ± 1.14 |

Presented are the factors; the means and standard deviations ($M$ ± SD) of the responses (ordered from highest mean to lowest); the results of a Welch two-sample $t$-test for differences between the genders in their responses, including the $t$-value (t), degrees of freedom (df) and the $p$-value ($p$); and the means and standard deviations of the respondents who identified as women and men.

It is possible, however, that people are not aware of factors that might actually be helpful. In order to uncover factors that could potentially be targeted to increase the number of women asking questions, we ran a series of multiple linear regressions on women survey respondents only, predicting how often they reported asking questions. First, we ran a model in which we entered (simultaneously) the five motivations for asking questions (see Supplementary Material 1, Q6). Three motivations predicted women reporting to ask more questions: being interested in the subject, $\beta$ = 0.14, $t(297) = 2.45$, $p = 0.02$; desiring clarification, $\beta$ = 0.12, $t(297) = 2.13$, $p = 0.03$; and wanting to act as a model for more junior academics, $\beta$ = 0.26, $t(297) = 4.54$, $p < 0.001$. Next, we tested a second model, in which we entered (simultaneously) three factors that are under departments' control: how much time is provided for questions, how many people usually attend, and how easy it is to meet the invited speakers informally. The only factor that predicted women reporting to ask more questions was fewer people attending the seminar, $\beta$ = -0.25, $t(297) = 4.36$, $p < 0.001$. Finally, we tested the extent to which the proportion of women faculty and women postgraduates (entered simultaneously) predicted women reporting to ask more questions, though this would obviously be difficult and time-consuming to change. Neither the proportion of women faculty, $\beta$ = 0.02, $t(292) = 0.26$, $p = 0.80$, nor the proportion of women postgraduates, $\beta$ = 0.06, $t(292) = 0.77$, $p = 0.44$, predicted women reporting that they would be more likely to ask questions.

Building on the findings from our pilot data that showed that the disparity decreased with a longer time dedicated to questions, we performed a manipulation that asked speakers to shorten their talks by 5-10 min in an effort to increase the proportion of questions from women. Over the two departments involved in the experiment, we collected data during 30 seminars, including 17 controls and 13 experimental seminars. Our manipulation was not successful: the duration of the time for questions was no longer in our treatment group than in the control group. In fact, there was a significant interaction between the institution and the treatment; on average, the time dedicated to questions increased in one institution from 14 to 16 min, but *decreased* in the other institution from 10 to 7 min. However, when considering the departments separately, these changes were not significant ($p > 0.05$). Thus, unsurprisingly, our manipulation did not have an effect on the proportion questions asked by women (generalised linear model with experimental condition as the only predictor: $\beta \pm$ S.E. = 0.30 ± 0.33, $z = 0.90$, $p = 0.37$).

To further explore whether it is possible to manipulate the time dedicated to questions to increase the gender balance in the questions asked, we ran a *post hoc* linear model investigating whether shorter seminars led to longer question periods using our full sample of observational data. Surprisingly, across all observed seminars, we found no association between the length of the seminar and the length of the question time (generalised linear model with a Poisson error structure for count data with duration of talk as a predictor of duration of question time: $\beta \pm$ S.E. = -0.002 ± 0.002, $z = -1.13$, $p = 0.26$). This result suggests that manipulating the talk duration would not result in a change in the time dedicated to questions. Therefore, the manipulation may have been more successful had we aimed to manipulate directly the time dedicated to questions rather than indirectly trying to affect this by manipulating the talk duration.

**Discussion**

The visibility of women role models at all career stages is important for redressing problems of the leaky pipeline. Our results add to a growing body of evidence showing that women are less visible than men, both conceptually and literally, in various scientific domains. Other studies have reported a similar bias in visibility, with men participating more already in school classrooms [23,27,28], at conferences [21,29], and public events [30]. Here, we report an underrepresentation in the literal visibility of women in a new domain: asking questions at departmental seminars. Our data show that a given question after a departmental seminar was more than 2.5 times more likely to be asked by a male than a female audience member, significantly misrepresenting the gender-ratio of the audience which was, on average, equal. These results are important because this gender disparity is observable particularly early in the career pipeline: junior academics are likely to observe the question-asking behaviour of the men and women in their department before they ever attend a conference, or become familiar with the researchers publishing in their area of interest. Below we briefly discuss the implications of our findings for women's attrition in academia, before addressing some limitations of our study and recommendations for increasing women's visibility at these events.

The lack of visible female role models asking questions at departmental seminars is likely to be both a symptom of the leaky pipeline *and* a cause of that same problem. As we explained earlier in this paper, research on role modelling suggests that having access to successful ingroup role models (e.g., women in senior levels of the academy) can be a key factor in determining what course of study or occupation a person will pursue [6,9], and, when people do not have first-hand experience in a particular domain, ingroup role models can signal whether a person would also be likely to achieve success in that domain [7,8]. In the case of academic seminars, then, the fact that our data show women asking disproportionately fewer questions than men necessarily means that junior scholars are encountering fewer visible female role models in the field. This lack of visibility of women during this type of regular academic interaction (the departmental seminar) is further compounded by women giving fewer talks at, and asking fewer questions at conferences [16,18,19], and women being less visible in the scientific literature as first and senior authors of scientific papers [10–13]. Given the importance of successful ingroup role modelling, we maintain that examining the visibility of female academics at local, departmental seminars is perhaps even more valuable than examining women's visibility at later levels of the academic trajectory (e.g., publications or conference presentations) because junior scholars are much more likely to attend these departmental seminars, as a way of "seeing what it is like" in order to make the choice of whether to pursue an academic career. Following from social role theory, from early on in their academic trajectory, scholars may encode the relative lack of female role models as an indicator that the academy is not a place where women are successful or represented, and subsequently choose to opt-out of academic careers as a result. When this happens, it perpetuates the original problem of the leaky pipeline by causing women who might have otherwise advanced to senior level positions in academia to take alternate career paths, which means there will continue to be fewer women than men in those positions.

One possible alternative interpretation of the low proportion of questions asked by women in our observational data is that more senior audience members are more willing to ask questions after seminars, and the data could accurately reflect the gender discrepancy in the proportions of senior audience members. That is, there could be a confound between seniority and gender, and the effect we observe is an effect of seniority, not of gender. Because we did not expect our observers to be familiar with the seniority of the members of the audience of all of the seminars they attended, we did not collect data on the seniority of the attendees asking questions. However, two lines of evidence suggest that the disparity we observed is not due only to this. First, in our observational data we controlled for the proportion of female faculty members in the host department and, while this proportion significantly predicted the proportion of questions asked by women, variation remained that was explained by other factors in the models. Additionally, this effect was "shallower" than a direct relationship would predict, with a 5% increase in the proportion of women faculty predicting only a 1.5% increase in the proportion of questions from women. This may suggest that senior women asked proportionally fewer questions than their senior male counter-parts, which is supported by our second line of evidence from the survey data. Men self-reported asking questions after seminars at higher frequencies than women at every career stage, suggesting that even amongst senior faculty men ask questions after seminars more frequently than women (Supplementary Material 4, Fig S4.1C). This finding is also consistent with one study of question-asking behaviour at conferences, which found that younger male attendees asked more questions than younger female attendees at the same rate as the entire sample of questions asked [21].Together, these patterns suggest that seniority does not completely explain the pattern we observed in the gender disparity.

Our observational data suggested that, in addition to the proportion of women faculty mentioned above, several factors affected the proportion of women asking questions after seminars. The proportion of women in the audience had a significant positive correlation with the proportion of questions asked by women. Although this result is unsurprising, the magnitude of the effect was relatively small, with only a ~1.6% higher share of questions asked by women for a 5% increase in women in the audience. Based on the results of the survey that showed that women rated internal factors as more important in preventing them from asking a question than men, we suggest that the weakness of this effect may stem from women's lower self-reported confidence when asking questions. Such an interpretation is further supported by the finding that a greater proportion of women asked questions when the speaker was from the department, suggesting that familiarity with the speaker may make asking a question less intimidating.

Contrary to our prediction, we found that when the speaker was male, a greater proportion of questions asked after the seminar were from women. We had predicted that the proportion of questions from women would be higher when the speaker was female. However, our results suggest that this was not the case and that men ask proportionally more questions of female speakers and/or women ask proportionally more questions of male speakers. One interpretation may be that men are less intimidated by female speakers than women are, and thus ask more questions when the speaker is female. Alternatively, or in addition to this interpretation, women may avoid "challenging" a female speaker, but may be less concerned for a male speaker.

The gender of the first person to ask a question was also correlated with the proportion of questions asked by women, with a greater proportion of women asking subsequent questions when the first question was asked by a woman compared to when the first question was asked by a man. A similar effect has also been observed at astronomy conferences [29]. We had included the gender of the first person to ask a question as a predictor because we believed that it may "set the tone" for subsequent questions. Our results suggest that this could be the case and may be an example of gender stereotype activation—where an individual behaves in a gender-stereotype consistent manner when a gender stereotype is activated [31,32]—with a male-first question *immediately* reinforcing gender stereotypes. This could affect not only women's but also men's behaviour after seminars, with women asking fewer questions and men asking more because of gender stereotypes in assertiveness and confidence. Alternatively, this association could arise because aspects we did not measure might have set an overall environmental tone influencing women and men to ask questions, with the first question being representative of any systematic bias in the subsequent questions. For example, it could be that because of internal factors women are only willing to ask questions in particularly stimulating situations, and in these situations, women will be both more likely to ask the first question, and to ask a greater-than-average proportion of questions. These alternative hypotheses result in the same prediction; an experimental approach is needed to tease them apart.

Several of these interpretations make connections between the self-report results, which focus on psychological factors, and the observational results, which focus on contextual factors. For example, we suggest that women's self-reported lower confidence might explain why they ask more questions when the speaker is internal. It is important to note that, despite research showing that people generally know their own personality best [33], they may lack self-knowledge [34,35], be inaccurate [36]. or may not wish to reveal their true feelings. On one hand, men's ratings of self-confidence may be low simply because they do not wish to report that they lack confidence. On the other hand, women may also not want to confirm stereotypes by reporting that they lack confidence, and their self-reports might be higher than reality. Thus, any comments on connections between the self-report and observational data are necessarily speculative.

*Some Recommendations*

Given the problem of the leaky pipeline and the importance of the visibility of women for addressing this, we hope to provide some recommendations that could increase women's visibility during these common events. First, however, we would like to make it clear that we do not place any blame on any party for the disparity that we observed in question-asking after seminars. Many men are not aware that men are asking proportionally more questions and most women identify internal factors as holding them back from asking questions. To the extent that participants' self-reports are accurate, our results suggest that internalised gender stereotypes may be at least partly responsible for the observed disparity [37] , both in men's participation and women's lack of it, and the problem can only be addressed by lasting changes in the academic culture that can help to break gender stereotypes and provide an environment which anyone can feel part of. However, until that time, our data suggest ways we could encourage more equal visibility of men and women, although we note that these recommendations have not yet been empirically tested.

Several of the factors that we identified as important to the proportion of women asking questions after seminars are not easily under a department's control, and we therefore do not consider them as actionable recommendations. These include the proportion of women in the audience and the proportion of women in the department, the latter of which could be changed only over the longer-term. While the characteristics of the speaker are difficult to manipulate, we would encourage seminar organisers not to neglect inviting internal speakers and for moderators to be particularly conscious of bias when the speaker is female. However, it may be possible to change the number of questions asked and the gender of the first person to ask a question. Increasing the number of questions increased the proportion of questions asked by women. Given our manipulation, which failed to increase question time by decreasing the seminar duration, we recommend that, where possible, the question time be unlimited, to encourage more questions. This could be achieved through, for example, booking a seminar room for longer than one hour so that the next event in the room does not cut short the question time. Having said this, a longer time for questions may be a taxing requirement for the speaker after having given a seminar, and may also be undesirable to the audience members. Alternatively, keeping questions and answers short (e.g., through an explicit statement of this as the new department culture, or with the help of a skilled moderator) will allow more questions to be asked during a given question period, and could be an alternative method to allow a greater proportion of questions from women. Although we cannot be sure of the causative relationship between the gender of the first questioner on the disparity in the subsequent questions asked, we would recommend that, should the opportunity arise, a female-first question be prioritised. This is because (a) a female-first question was a good predictor of low disparity in the questions asked in our observational data and it is possible that gender stereotype activation is responsible for the observed difference and (b) by choosing a female-first question, a female-friendly environment may be fostered over time.

Generally, we feel that more could be done through active changes in speakers', attendees' and particularly moderators' behaviour. Having an active, trained moderator may avoid those situations where one audience member seems to be "showing off" (which survey respondents claim to be the case quite often; $N_{male}$ = 9, $N_{female}$ = 10) or is going off-topic, or a speaker who goes over time. In addition, moderators could be trained to see the whole room (location was mentioned as a factor by $N_{male}$ = 2, $N_{female}$ = 2), and to maintain as much balance as possible with respect to gender and seniority of question-askers. In the open-ended survey questions, respondents complained that moderators call on people they know or more senior people, overlooking the rest ($N_{male}$ = 3, $N_{female}$ = 6). Although it may seem fair to call on people in the order that they raise their hands, doing so may inadvertently result in fewer women and junior academics asking questions, since they often need more time to formulate questions and work up the nerve. In our observational data, we did not record whether a moderator was present, and we did not record the gender of people who attempted to ask a question; our data cannot elucidate whether women asked fewer questions because fewer women raised their hand or because fewer women were chosen to ask a question. It is likely that the discrepancy results from a combination of both, supporting the potential benefits of an active moderator.

Women rated internal factors as more important in holding them back from asking a question, compared to men. To counteract this low confidence, it could help both women and men to provide a small break between the talk and the question period, which would give people time to formulate a question and try it out on a colleague, as well as providing the general benefits of allowing people who need to leave a chance to do so, and giving the speaker a small break.

Although these recommendations (which have yet to be tested empirically) were generated with the idea of increasing women's visibility, they are likely to benefit everyone. It is not only women who are underrepresented in academia; aspiring and early career academics would also benefit from ethnic minorities being more visible.

Our results have implications for redressing the leaky pipeline in academia and indicate that without active steps, the various factors that contribute to women choosing other careers over academia are unlikely to change. Our results support a self-perpetuating feedback loop, where the absence of visible role models influences the behaviour of women in a way that is likely to increase their decision to leave, further reducing their visibility. However, our data show that women are not inherently less likely to ask questions when the conditions are favourable—there is no gender discrepancy when a woman asks the first question. Our suggestions should be seen as aims to create favourable conditions that remove the barriers that restrain anyone from speaking up and being visible.

**Supporting Information**

S1 Qualtrics_Survey_Participation_in_seminars. **Survey questions and survey flow.**

Caption: The list of questions and the flow of the Qualtrics survey

S2 methodology for collecting data in seminars. **Instructions for collecting observational data.**

Caption: The instructions that were distributed to the observers who agreed to collect data at their local seminars.

S3 Detailed Results Tables. **Detailed breakdowns of the survey data by gender**

Caption: The tables show a more detailed breakdown of the survey data by gender than what is reported in the text. Included are tables that break down the answers to the questions about (1) why respondents sometimes do not ask questions when they have one; (2) if they believe that women ask fewer questions, why they think that women do not ask questions; and (3) factors they think may be important in encouraging women to ask questions after academic seminars.

S4 seniority and gender. **The effect of seniority and gender on self-reported question-asking behaviour in seminars**

Caption: The results of linear mixed models are presented that test for relationships between survey respondents' gender and seniority on their self-reported frequency of asking questions. A figure is also included that reflect these data.

**Acknowledgements**

We have compiled a list of resources about various gender biases in academia (e.g., career progression, authorship, funding, awards…) on our project website: http://diversityinacademia.strikingly.com/. We welcome your help in building and maintaining this list, to benefit anyone who is interested in learning more about these issues. We thank the following people for generously attending seminars near them, counting people and questions, and staying until the very end, even if it took 60 min (listed alphabetically by first name): Aaron Weidman, Adrian Currie, Alan McElligot, Alexis May, Anne Charmantier, Anne Goldizen, Anne Pisor, Caroline Spence, Christine Webb, Corina Logan, Culum Brown, Danny Osborne, Dominique Knutsen, Elva Robinson, Federica Dal Pesco, Gabrielle Davidson, Geoff While, Gisela Kopp, Greg Brooke, Guy Cowlishaw, Harry Marshall, Helge Gillmeister, James Gilbert, Joey Cheng, Kate Laskowski, Kirsty MacLeod, Lara Aknin, Lauren Brent, Lauren Human, Matteo Rizzuto, Melanie Dahmhann, Naomi Langmore, Pierre-Olivier Montiglio, Sandra Binning, Sandrine Muller, Sharon Morein, Sonja Koski, Stefan Fischer, Stuart Semple, Teri Kirby, Theadora Block, and Wendy King. We also thank the reviewers and everyone who provided feedback on the preprint (https://arxiv.org/abs/1711.10985), which greatly improved the manuscript. AJC was supported by a Junior Research Fellowship from Churchill College, University of Cambridge during the conception of the study and collection of data.

## References

1. Ceci SJ, Ginther DK, Kahn S, Williams WM. Women in Academic Science: A Changing Landscape. Psychol Sci Public Interest. 2014;15: 75–141. doi:10.1177/1529100614541236

2. Alper J. The Pipeline Is Leaking Women All the Way Along. Science. 1993;260: 409. doi:10.1126/science.260.5106.409

3. Damschen EI, Rosenfeld KM, Wyer M, Murphy-Medley D, Wentworth TR, Haddad NM. Visibility matters: increasing knowledge of women's contributions to ecology. Front Ecol Environ. 2005;3: 212–219.

4. Eagly AH, Steffen VJ. Gender stereotypes stem from the distribution of women and men into social roles. J Pers Soc Psychol. 1984;46: 735.

5. Peters K, Ryan M, Haslam SA, Fernandes H. To Belong or Not to Belong. J Pers Psychol. 2012;

6. Bandura A. Self-Efficacy: The Exercise of Control. 1st ed. New York: W. H. Freeman; 1997.

7. Schmader T, Major B. The impact of ingroup vs outgroup performance on personal values. J Exp Soc Psychol. 1999;35: 47–67.

8. Tindale RS, Kulik CT, Scott LA. Individual and group feedback and performance: An attributional perspective. Basic Appl Soc Psychol. 1991;12: 41–62.

9. Stout JG, Dasgupta N, Hunsinger M, McManus MA. STEMing the tide: using ingroup experts to inoculate women's self-concept in science, technology, engineering, and mathematics (STEM). J Pers Soc Psychol. 2011;100: 255.

10. Brown AJ, Goh JX. Some evidence for a gender gap in personality and social psychology. Soc Psychol Personal Sci. 2016;7: 437–443.

11. Filardo G, da Graca B, Sass DM, Pollock BD, Smith EB, Martinez MA-M. Trends and comparison of female first authorship in high impact medical journals: observational study (1994-2014). bmj. 2016;352: i847.

12. Jagsi R, Guancial EA, Worobey CC, Henault LE, Chang Y, Starr R, et al. The "gender gap" in authorship of academic medical literature—a 35-year perspective. N Engl J Med. 2006;355: 281–287.

13. West JD, Jacquet J, King MM, Correll SJ, Bergstrom CT. The role of gender in scholarly authorship. PloS One. 2013;8: e66212.

14. Maliniak D, Powers R, Walter BF. The gender citation gap in international relations. Int Organ. 2013;67: 889–922.

15. Macaluso B, Larivière V, Sugimoto T, Sugimoto CR. Is science built on the shoulders of women? A study of gender differences in contributorship. Acad Med. 2016;91: 1136–1142.

16. Isbell LA, Young TP, Harcourt AH. Stag Parties Linger: Continued Gender Bias in a Female-Rich Scientific Discipline. PLOS ONE. 2012;7: e49682. doi:10.1371/journal.pone.0049682

17. Johnson CS, Smith PK, Wang C. Sage on the Stage: Women's Representation at an Academic Conference. Pers Soc Psychol Bull. 2017;43: 493–507.

18. Schroeder J, Dugdale HL, Radersma R, Hinsch M, Buehler DM, Saul J, et al. Fewer invited talks by women in evolutionary biology symposia. J Evol Biol. 2013;26: 2063–2069. doi:10.1111/jeb.12198

19. Jones TM, Fanson KV, Lanfear R, Symonds MRE, Higgie M. Gender differences in conference presentations: a consequence of self-selection? Stewart G, editor. PeerJ. 2014;2: e627. doi:10.7717/peerj.627

20. Davenport JR, Fouesneau M, Grand E, Hagen A, Poppenhaeger K, Watkins LL. Who asks questions at astronomy meetings? ArXiv Prepr ArXiv14033091. 2014;

21. Hinsley A, Sutherland WJ, Johnston A. Men ask more questions than women at a scientific conference. PloS One. 2017;12: e0185534.

22. Pritchard J, Masters K, Allen J, Contenta F, Huckvale L, Wilkins S, et al. Asking gender questions. Astron Geophys. 2014;55: 6–8.

23. Eddy SL, Brownell SE, Wenderoth MP. Gender gaps in achievement and participation in multiple introductory biology classrooms. CBE-Life Sci Educ. 2014;13: 478–492.

24. Kelly A. Gender differences in teacher–pupil interactions: a meta-analytic review. Res Educ. 1988;39: 1–23. doi:10.1177/003452378803900101

25. Bates D, Mächler M, Bolker B, Walker S. Fitting Linear Mixed-Effects Models Using lme4. J Stat Softw. 2015;67: 1–48. doi:10.18637/jss.v067.i01

26. Muchnik L, Aral S, Taylor SJ. Social influence bias: A randomized experiment. Science. 2013;341: 647–651.

27. Alexander CS, others. Study on Women's Experiences at Harvard Law School. 2004;

28. Yale Law Women. Yale Law School Faculty and Students Speak up about Gender: Ten Years Later. Yale Law Women; 2012.

29. Schmidt SJ, Davenport JR. Who asks questions at astronomy meetings? Nature Astronomy 1: 0153. doi:10.1038/s41550-017-0153

30. Holmes J. Women's talk in public contexts. Discourse Soc. 1992;3: 131–150.

31. Wheeler SC, Petty RE. The effects of stereotype activation on behavior: a review of possible mechanisms. Psychol Bull. 2001;127: 797.

32. Banaji MR, Hardin CD. Automatic Stereotyping. Psychol Sci. 1996;7: 136–141. doi:10.1111/j.1467-9280.1996.tb00346.x

33. Funder DC, Colvin CR. Congruence of others' and self-judgments of personality. In: Hogan R, Johnson J, Briggs S, editors. Handbook of personality psychology. San Diego: Academic Press; 1997 pp. 617-647.

34. John OP, Robins RW. Determinants of interjudge agreement on personality traits: The Big Five domains, observability, evaluativeness, and the unique perspective of the self. J Pers. 1993; 61(4):521-551.

35. Vazire S. Who knows what about a person? The self–other knowledge asymmetry (SOKA) model. J Pers Soc Psychol. 2010;98(2):281-300.

36. Dunning D, Heath C, Suls JM. Flawed self-assessment: Implications for health, education, and the workplace. Psychol Sci Public Interest. 2004;5(3):69-106.

37. McClean E, Martin SR, Emich KJ, Woodruff T. The social consequences of voice: An examination of voice type and gender on status and subsequent leader emergence. Acad Manage J. 2017;

Figure1

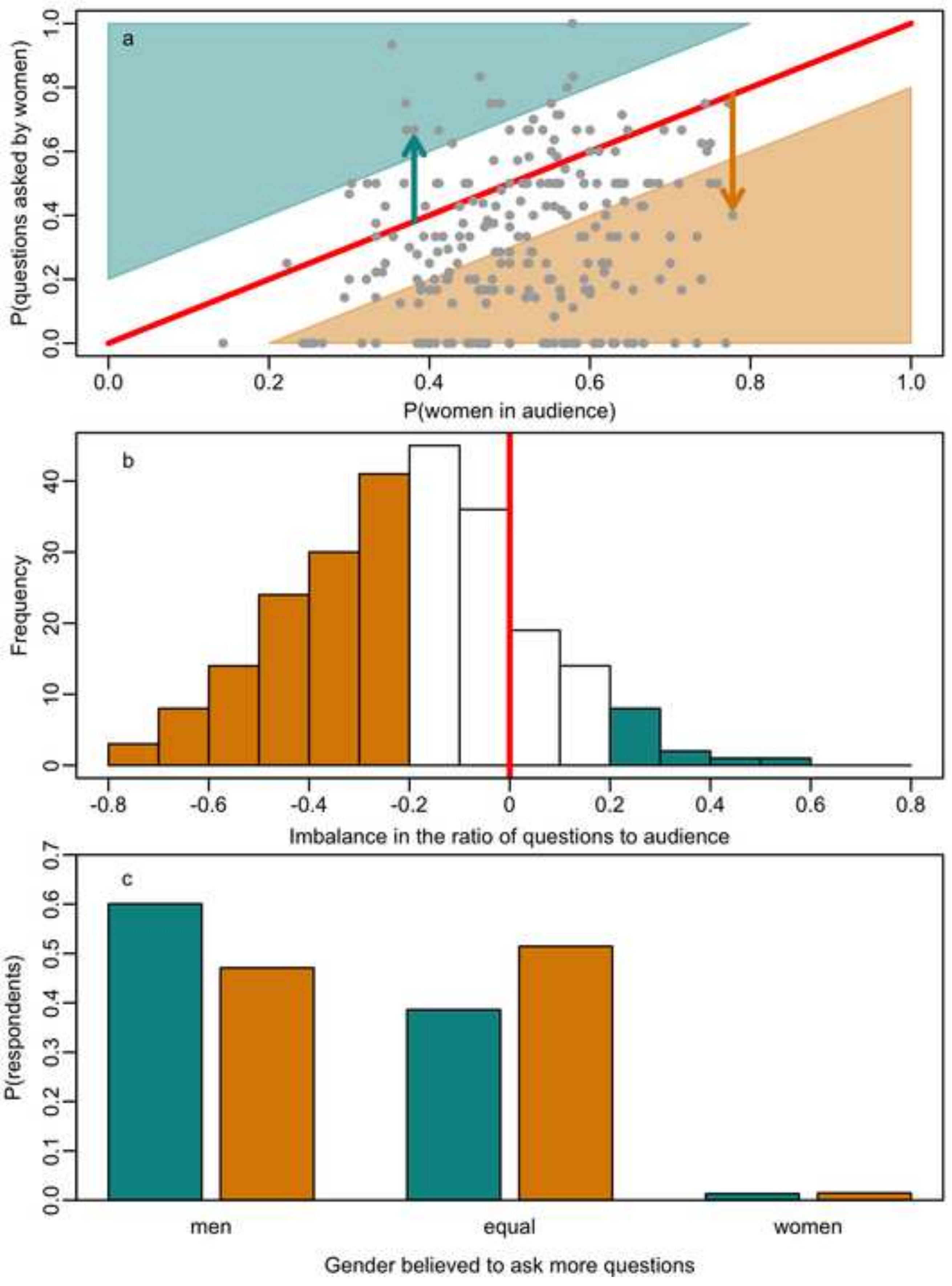
a
P(questions asked by women)
0.0
0.2
0.4
0.6
0.8
1.0
P(women in audience)
b
Frequency
0
10
20
30
40
-0.8
-0.6
-0.4
-0.2
0
0.2
0.4
0.6
0.8
Imbalance in the ratio of questions to audience
c
P(respondents)
0.0
0.1
0.2
0.3
0.4
0.5
0.6
0.7
men
equal
women
Gender believed to ask more questions

Figure2

Click here to download Figure Fig2.tif

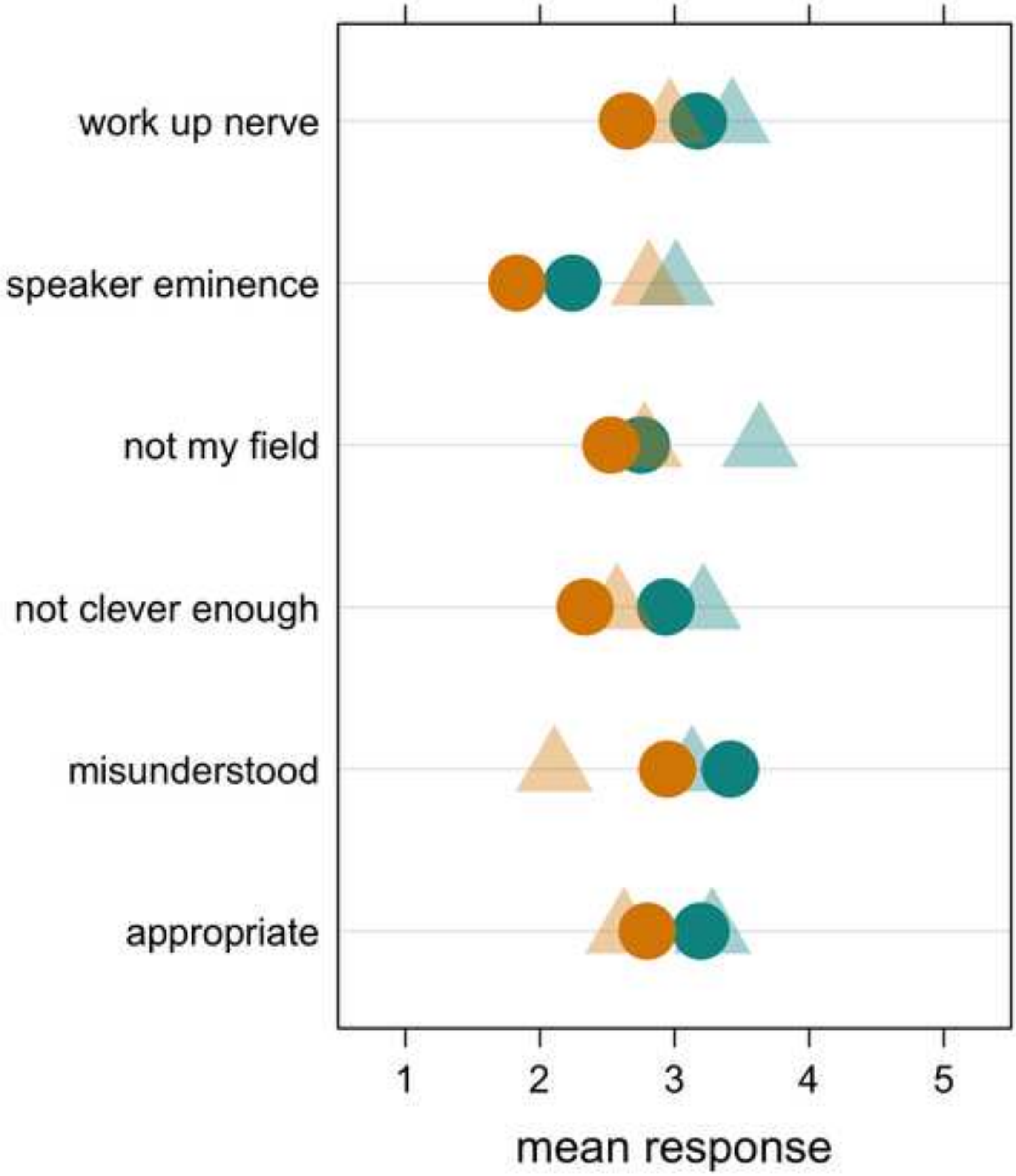

Figure3

Click here to download Figure Fig3.tif

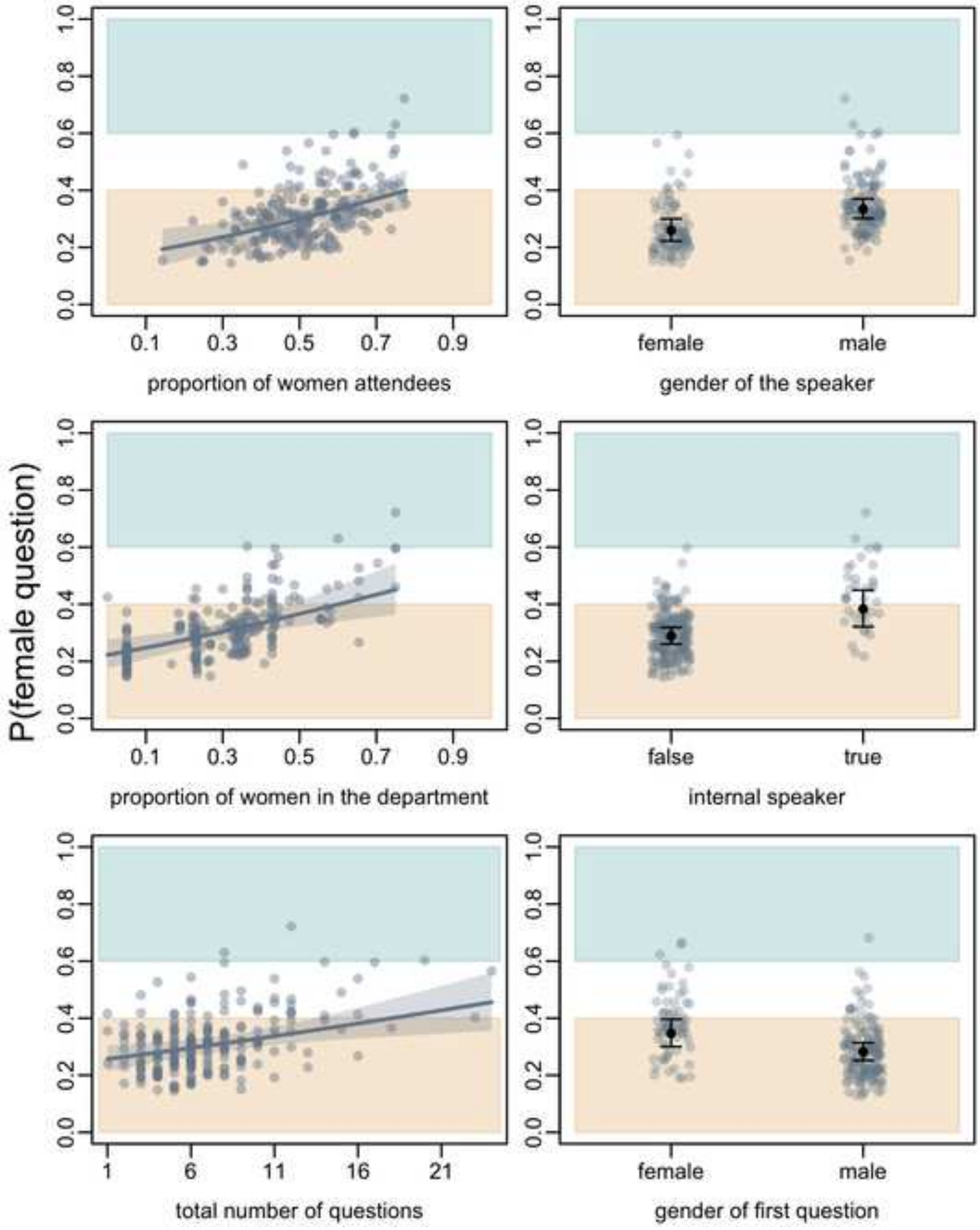

Participation in seminars

Q1 Thank you for taking the time to participate! In this study, you will be asked some questions about your attendance and participation in academic seminars and the culture around participation in seminars in your department, before we ask some questions about your demography. The study will take fewer than 10 minutes to complete. You can withdraw your participation at any time during the study and are never obliged to answer a question. Your privacy is very important. We will ask you for some demographic information, but nothing that could identify you. We are committed to open science, so the data that we collect will be made available online, for use by other researchers (e.g., at https://osf.io/). Results from this study will be presented in academic publications. Results are normally presented in terms of groups of individuals. If any individual data were presented, the data would be anonymous, without any means of identifying the individuals involved. This project has received ethical approval from the Departmental Psychology Ethics Committee of the University of Essex. If you have any questions you'd like to ask before starting the survey, please feel free to contact Dr. Gillian Sandstrom at gsands@essex.ac.uk, Dr Alecia Carter at ac854@cam.ac.uk, Dr Dieter Lukas at dl384@cam.ac.uk or Dr Alyssa Croft at alyssac@email.arizona.edu. Consent to Participate: I have read and understood the consent form. I have had sufficient time to consider the information provided and to ask for advice if necessary. I have had the opportunity to ask questions and have had satisfactory responses to my questions. I understand that all of the information collected will be kept confidential and that the results will be made publically available. I understand that my participation in this study is voluntary and that I am completely free to refuse to participate or to withdraw from this study at any time. I understand that I am not waiving any of my legal rights as a result of agreeing to this consent form. If you consent, please click the 'I consent' button below and then click the arrow to begin the survey.

- ❍ I consent (1)

If I consent Is Not Selected, Then Skip To End of Survey

Q2 For this survey, we define a seminar as a public presentation at an academic institution attended by students and faculty. On average, how many of these types of seminars have you attended?

- ❍ >1 per week (1)
- ❍ Weekly (2)
- ❍ Fortnightly/Bi-weekly (3)
- ❍ Monthly (4)
- ❍ A few times per year (5)

Q3 Do you ask questions in:

- ❍ All seminars? (1)
- ❍ Most seminars? (2)
- ❍ Some seminars? (3)
- ❍ Few seminars? (4)
- ❍ No seminars? (5)

Q4 Have you ever NOT asked a question when you wanted to?

❍ Yes (sometimes I don't ask questions, even when I have one) (1)

❍ No (I always ask the questions that I want to ask) (2)

If No (I always ask the questi... Is Selected, Then Skip To If you ask questions at seminars, wha...

Q5 How important is each of these factors in stopping you from asking a question?

| | Not at all important (1) | Slightly important (2) | Moderately important (3) | Very important (4) | Extremely important (5) |
|---|---|---|---|---|---|
| Not enough time (1) | ❍ | ❍ | ❍ | ❍ | ❍ |
| Worried that I had misunderstood the content (2) | ❍ | ❍ | ❍ | ❍ | ❍ |
| Couldn't work up the nerve (3) | ❍ | ❍ | ❍ | ❍ | ❍ |
| Not sure whether the question was appropriate (4) | ❍ | ❍ | ❍ | ❍ | ❍ |
| Not my field (5) | ❍ | ❍ | ❍ | ❍ | ❍ |
| The speaker was too eminent/intimidating (6) | ❍ | ❍ | ❍ | ❍ | ❍ |
| Worried that I was not clever enough to ask a good question (7) | ❍ | ❍ | ❍ | ❍ | ❍ |
| I was meeting the speaker later / asked after the talk had ended (8) | ❍ | ❍ | ❍ | ❍ | ❍ |
| Other (please specify): (9) | ❍ | ❍ | ❍ | ❍ | ❍ |

Q7 To what extent would each of these factors encourage you to ask more questions?

| | Wouldn't help at all (1) | Wouldn't help much (2) | Might help a bit (3) | Would help a lot (4) | Would make a huge difference (5) |
|---|---|---|---|---|---|
| A longer time to formulate the question (1) | ❍ | ❍ | ❍ | ❍ | ❍ |
| A chance to ask in person (2) | ❍ | ❍ | ❍ | ❍ | ❍ |
| Nicer speakers (3) | ❍ | ❍ | ❍ | ❍ | ❍ |
| More welcoming host (4) | ❍ | ❍ | ❍ | ❍ | ❍ |
| Confidence (5) | ❍ | ❍ | ❍ | ❍ | ❍ |
| Seniority (6) | ❍ | ❍ | ❍ | ❍ | ❍ |
| Having a moderator to ask the questions (7) | ❍ | ❍ | ❍ | ❍ | ❍ |
| Moderator doing a better job engaging whole audience (8) | ❍ | ❍ | ❍ | ❍ | ❍ |
| Other (please specify): (9) | ❍ | ❍ | ❍ | ❍ | ❍ |

Q6 If / when you ask questions at seminars, what has been your main motivation? (Check all that apply.)

❑ Interested in subject (1)
❑ Feel you spotted a mistake (2)
❑ Need for clarification (3)
❑ I feel it's part of my role (e.g., to act as a model for more junior academics) (4)
❑ To establish a connection with a particular speaker (5)

Q28 What factors do you think play a role in who asks questions?Does Seniority play a role in who asks questions in seminars?

- ❍ Senior audience members ask more questions (1)
- ❍ Junior audience members ask more questions (2)
- ❍ Senior and Junior audience members ask about the same amount of questions (3)

Q30 Does Confidence play a role in who asks questions in seminars?

- ❍ Confident people ask more questions (1)
- ❍ Not confident people ask more questions (2)
- ❍ Confident and Not confident people ask about the same amount of questions (3)

Q27 Does Extraversion play a role in who asks questions in seminars?

- ❍ Introverted people ask more questions (1)
- ❍ Extraverted people ask more questions (2)
- ❍ Introverted and Extraverted people ask about the same amount of questions (3)

Q29 Does Gender play a role in who asks questions in seminars?

- ❍ Women ask more questions (1)
- ❍ Men ask more questions (2)
- ❍ Men and Women ask about the same amount of questions (3)

Q31 Does Competence play a role in who asks questions in seminars?

- ❍ Competent people ask more questions (1)
- ❍ Incompetent people ask more questions (2)
- ❍ Competent and Incompetent people ask about the same amount of questions (3)

Q28 Do other factors play a role in who asks questions during seminars?

- ❍ No (4)
- ❍ Maybe (5) ____________________
- ❍ Yes (6) ____________________

Answer If Does Gender play a role in who asks questions in seminars? Men ask more questions Is Selected Or Does Gender play a role in who asks questions in seminars? Women ask more questions Is Selected

Q34 You indicated that gender plays a role in who asks questions. How important do you think each of these factors is in preventing the gender asking fewer questions from asking more questions?

| | Not at all important (1) | Slightly important (2) | Moderately important (3) | Very important (4) | Extremely important (5) |
|---|---|---|---|---|---|
| Worry that they misunderstand the content (1) | ❍ | ❍ | ❍ | ❍ | ❍ |
| Can't work up the nerve (2) | ❍ | ❍ | ❍ | ❍ | ❍ |
| Are unsure that their questions are appropriate (3) | ❍ | ❍ | ❍ | ❍ | ❍ |
| Feel they are not an expert (4) | ❍ | ❍ | ❍ | ❍ | ❍ |
| Feel intimidated by the speaker (5) | ❍ | ❍ | ❍ | ❍ | ❍ |
| Believe that they are not clever enough to ask a good question (6) | ❍ | ❍ | ❍ | ❍ | ❍ |
| Ask questions after the seminar is over (7) | ❍ | ❍ | ❍ | ❍ | ❍ |
| Other (please specify): (8) | ❍ | ❍ | ❍ | ❍ | ❍ |

Q14 How much time is usually provided for questions after seminars?

❍ (1)
❍ 5-10 min (2)
❍ 11-15 min (3)
❍ 16-30 min (4)
❍ >30 min (5)

Q15 How many people usually attend your departmental seminars?

- (1)
- 15-30 (2)
- 31-45 (3)
- >45 (4)

Q27 How easy is it to meet invited speakers informally?

- Extremely easy (1)
- Easy enough (2)
- Not easy or difficult (3)
- A bit difficult (4)
- It's not possible (5)

Q17 What is the culture around meeting speakers in your department?

- Speakers meet only with the host (1)
- Speakers meet with relevant faculty (2)
- Anyone can sign up to meet a speaker (3)
- Everyone is actively encouraged to meet speakers, and speakers have organized events (e.g. lunch with PhD students) (4)

Q13 In your department, what percentage of the permanent faculty are women?

- (1)
- 10-25% (2)
- 26-50% (3)
- 51-75% (4)
- 76-100% (5)
- I don't know (6)

Q28 In your department, what percentage of the graduate/PhD students are women?

- (1)
- 10-25% (2)
- 26-50% (3)
- 51-75% (4)
- 76-100% (5)
- I don't know (6)

Q11 What is your subject? (Please choose one from the list.)

- Accounting and Finance (1)
- Anthropology (2)
- Archaeology (3)
- Art History (4)
- Biochemistry (5)
- Biological Sciences (6)
- Biomedical Sciences (7)
- Business and Management (8)
- Chemistry (9)
- Classics and Ancient History (10)
- Computer Science and IT (11)
- Criminology (12)
- Drama (13)
- Earth Sciences (14)
- Economics (15)
- Education (16)
- Engineering (17)
- English language (18)
- English literature (19)
- Environmental Science (20)
- Film Studies (21)
- Geography (22)
- Geology (23)
- History (24)
- Human Sciences (25)
- International studies (26)
- Law (27)
- Liberal Arts (28)
- Linguistics (29)
- Materials Science (30)
- Mathematics (31)
- Medicine (32)
- Modern Languages (33)
- Music (34)
- Natural Sciences (35)
- Neuroscience (36)
- Philosophy (37)
- Physics and Astronomy (38)
- Politics and International Relations (39)
- Psychology (40)
- Social Sciences (41)
- Sociology and Criminology (42)
- Sport and Health Sciences (43)

❍ Theology and Religion (44)
❍ Something else (45)

Q18 What career stage are you?
❍ Undergraduate student (1)
❍ Postgraduate student (e.g., Masters, PhD) (2)
❍ Postdoctoral researcher (3)
❍ Research fellow (4)
❍ Faculty (5)
❍ Other (please specify) (6) ____________________

Q19 How long have you been at this stage?
❍ (1)
❍ 1-2 years (2)
❍ 3-5 years (3)
❍ 6-10 years (4)
❍ >10 years (5)

Q21 In which country is your current institution?
- Afghanistan (1)
- Albania (2)
- Algeria (3)
- Andorra (4)
- Angola (5)
- Antigua and Barbuda (6)
- Argentina (7)
- Armenia (8)
- Australia (9)
- Austria (10)
- Azerbaijan (11)
- Bahamas (12)
- Bahrain (13)
- Bangladesh (14)
- Barbados (15)
- Belarus (16)
- Belgium (17)
- Belize (18)
- Benin (19)
- Bhutan (20)
- Bolivia (21)
- Bosnia and Herzegovina (22)
- Botswana (23)
- Brazil (24)
- Brunei Darussalam (25)
- Bulgaria (26)
- Burkina Faso (27)
- Burundi (28)
- Cambodia (29)
- Cameroon (30)
- Canada (31)
- Cape Verde (32)
- Central African Republic (33)
- Chad (34)
- Chile (35)
- China (36)
- Colombia (37)
- Comoros (38)
- Congo, Republic of the... (39)
- Costa Rica (40)
- Côte d'Ivoire (41)
- Croatia (42)
- Cuba (43)

- Cyprus (44)
- Czech Republic (45)
- Democratic People's Republic of Korea (46)
- Democratic Republic of the Congo (47)
- Denmark (48)
- Djibouti (49)
- Dominica (50)
- Dominican Republic (51)
- Ecuador (52)
- Egypt (53)
- El Salvador (54)
- Equatorial Guinea (55)
- Eritrea (56)
- Estonia (57)
- Ethiopia (58)
- Fiji (59)
- Finland (60)
- France (61)
- Gabon (62)
- Gambia (63)
- Georgia (64)
- Germany (65)
- Ghana (66)
- Greece (67)
- Grenada (68)
- Guatemala (69)
- Guinea (70)
- Guinea-Bissau (71)
- Guyana (72)
- Haiti (73)
- Honduras (74)
- Hong Kong (S.A.R.) (75)
- Hungary (76)
- Iceland (77)
- India (78)
- Indonesia (79)
- Iran, Islamic Republic of... (80)
- Iraq (81)
- Ireland (82)
- Israel (83)
- Italy (84)
- Jamaica (85)
- Japan (86)
- Jordan (87)

- Kazakhstan (88)
- Kenya (89)
- Kiribati (90)
- Kuwait (91)
- Kyrgyzstan (92)
- Lao People's Democratic Republic (93)
- Latvia (94)
- Lebanon (95)
- Lesotho (96)
- Liberia (97)
- Libyan Arab Jamahiriya (98)
- Liechtenstein (99)
- Lithuania (100)
- Luxembourg (101)
- Madagascar (102)
- Malawi (103)
- Malaysia (104)
- Maldives (105)
- Mali (106)
- Malta (107)
- Marshall Islands (108)
- Mauritania (109)
- Mauritius (110)
- Mexico (111)
- Micronesia, Federated States of... (112)
- Monaco (113)
- Mongolia (114)
- Montenegro (115)
- Morocco (116)
- Mozambique (117)
- Myanmar (118)
- Namibia (119)
- Nauru (120)
- Nepal (121)
- Netherlands (122)
- New Zealand (123)
- Nicaragua (124)
- Niger (125)
- Nigeria (126)
- Norway (127)
- Oman (128)
- Pakistan (129)
- Palau (130)
- Panama (131)

- Papua New Guinea (132)
- Paraguay (133)
- Peru (134)
- Philippines (135)
- Poland (136)
- Portugal (137)
- Qatar (138)
- Republic of Korea (139)
- Republic of Moldova (140)
- Romania (141)
- Russian Federation (142)
- Rwanda (143)
- Saint Kitts and Nevis (144)
- Saint Lucia (145)
- Saint Vincent and the Grenadines (146)
- Samoa (147)
- San Marino (148)
- Sao Tome and Principe (149)
- Saudi Arabia (150)
- Senegal (151)
- Serbia (152)
- Seychelles (153)
- Sierra Leone (154)
- Singapore (155)
- Slovakia (156)
- Slovenia (157)
- Solomon Islands (158)
- Somalia (159)
- South Africa (160)
- Spain (161)
- Sri Lanka (162)
- Sudan (163)
- Suriname (164)
- Swaziland (165)
- Sweden (166)
- Switzerland (167)
- Syrian Arab Republic (168)
- Tajikistan (169)
- Thailand (170)
- The former Yugoslav Republic of Macedonia (171)
- Timor-Leste (172)
- Togo (173)
- Tonga (174)
- Trinidad and Tobago (175)

- ❍ Tunisia (176)
- ❍ Turkey (177)
- ❍ Turkmenistan (178)
- ❍ Tuvalu (179)
- ❍ Uganda (180)
- ❍ Ukraine (181)
- ❍ United Arab Emirates (182)
- ❍ United Kingdom of Great Britain and Northern Ireland (183)
- ❍ United Republic of Tanzania (184)
- ❍ United States of America (185)
- ❍ Uruguay (186)
- ❍ Uzbekistan (187)
- ❍ Vanuatu (188)
- ❍ Venezuela, Bolivarian Republic of... (189)
- ❍ Viet Nam (190)
- ❍ Yemen (191)
- ❍ Zambia (192)
- ❍ Zimbabwe (193)

Q12 What is your gender?

- ❍ Male (1)
- ❍ Female (2)
- ❍ Transgender (3)
- ❍ Prefer not to say (4)

Q22 The study's aims: This study was designed to help us understand why there is a bias in the gender ratio of academics that attend and ask questions during seminars. Our preliminary research shows that more women attend seminars than men, but they ask fewer questions. From your answers, we would like to make recommendations that will lead to an improved visibility of women in academia through fostering an environment that promotes women's participation in regular academic events. The last thing that we want to ask you is not to share your knowledge about the true purpose of this study. We will be running this study for several weeks. As you can imagine, it would be very difficult for us to collect accurate information if people knew about the true purpose of this study beforehand. Consequently, we would appreciate if you do not discuss the true aim of this survey with others. Thank you so much for participating in this research. Without your help we would be unable to test our hypotheses and gather the necessary data. In case you are interested in the findings of the survey, we will be updating this website once the survey is completed http://academicseminarparticipation.strikingly.com/ If you have any questions, please contact any of the investigators on the project: Dr. Gillian Sandstrom (gsands@essex.ac.uk); Dr Alecia Carter (ac854@cam.ac.uk); Dr Dieter Lukas (dl384@cam.ac.uk); Dr Alyssa Croft (alyssac@email.arizona.edu).

# Women's visibility in academia: why don't women ask questions in seminars?

Drs Alecia Carter, Dieter Lukas, Gillian Sandstrom, Alyssa Croft

Brief background: The attrition of women in academia is a major concern, particularly in Science, Technology, Engineering, and Mathematics subjects. One factor that may contribute to the attrition of women is the lack of visibility of women in academia, which may result in women feeling like they do not belong. One place where academics regularly meet and see other academics is at departmental seminars. Our preliminary research suggests that more women attend departmental seminars than men, but ask disproportionally fewer questions than male attendees. However, thus far, we have collected data only at seminars in the University of Cambridge, which may not be representative of trends elsewhere. Our aim is to broaden our sample of seminars to other subjects, institutions, and countries to determine whether the trend is similar. We would be very grateful if you would help by collecting data in your department. We hope that this will not be onerous, as it will involve only a bit of counting during seminars you're already attending during one semester (and staying until the very end of the seminar!). As it would be better for our analyses if we could control for departmental differences using random effects, we would like a sample of about 3-4 seminars per department i.e. 3 or 4 data from you. This is not absolutely necessary, of course, but would be lovely to get, if possible. Below we describe the data we have been collecting, and that we will ask you to collect.

Methodology: To comply with ethics, data can be collected passively only at *public* events i.e. seminars that are advertised widely and can be attended by the public. Most departmental seminars fit into this category. Seminars within lab group probably do not count—please do not collect data at these.

We would like to ensure that our activities do not affect the behaviour of the audience (and affect the data we collect), ***so we please ask that you do not tell your colleagues about our study***.

The data we collect will be pooled into a google sheet "question_gender_groupsheet". Please click here to request access to a Google sheet for entering your data:

https://docs.google.com/spreadsheets/d/1pr2zj97e1hFHF-BqS49aHBX2csGS2r0infMlK49WdLY/edit?usp=sharing

The data to be collected during the talks are very straightforward, as is the google sheet. There is a greyed line of example data at the top of this sheet for guidance.

The column headings and data (* indicates data to be collected during the seminars, whereas all other data can be collected at another time) to be inputted are:

**country**: the country in which the talk took place

**university**: the university / institution in which the talk took place

**department**: the name of the department in which the talk took place. This is to control not only for differences between departments, but also so that we can retrospectively control for differences between subjects. We've been using short versions such as "Zoology" and "BioAnth" instead of "Department of Zoology" and "Department of Biological Anthropology".

**observer**: your name (will not be published! Just for queries)

**speaker**: the speaker's name (this information will not be shared, it is collected only so that we can collect further information on career stage, speakers' institutions etc. later if necessary)

**speakerInternal**: T or F. Is the speaker a member of your department? (To control for whether it is easier to ask questions of speakers with whom you are familiar.)

**speakerSex**: M or F for male or female, respectively

**title**: the title of the talk (again, so that we can retrospectively code the subject if necessary)

**date**: the date of the talk as DD/MM/YYYY

**time**: the time of the talk as HHMM e.g. 1630

***N_attendees**: number of attendees at the seminar. Includes the host (if any), and the observer (i.e. you), but not the speaker. Please be accurate—count twice (and more times if you have disagreement!). Many people arrive late i.e.

after the talk has started. We have been including these individuals as they are often obvious. However, we have NOT been reducing the totals for people who leave towards the end of the seminar.

***N_male**: the number of male attendees at the seminar. As for N_attendees, please count at least twice for either males or females or both.

***N_female**: the number of female attendees at the seminar.

*What we define as a question:* For the following data, there is a tricky <ahem> question of what constitutes an independent question. This is because the same attendee may ask two questions in a row, or an attendee may follow-up on/add to a question asked by another attendee, etc. As we are interested in visibility, a 'question' may be a comment or statement made by an attendee, rather than a question requiring an answer (though it usually does get a comment from the speaker). Follow-up questions (i.e. asking for *clarification* of the response of the speaker) are not included, but a subsequent question by the same attendee requiring a different response is included. We have thus defined a question as "*a statement made by an attendee that solicits a response from the speaker*". It is NOT the number of attendees who asked questions. It is thus feasible (but unlikely) that M/F_question could be greater than N_male / N_female. Questions that are asked during talks (i.e. the speaker is interrupted), are included, but are recorded in a separate column as we are interested in the gender of people who interrupt a talk (this is a rare occurrence for us, but may be more habitual elsewhere). Please include your own question should you ask one.

***First_question**: F or M. The gender of the attendee that asked the first question during the 'official' question period (i.e. after the talk has ended).

***M_question**: the number of questions asked by male attendees.

***F_question**: the number of questions asked by female attendees.

***M_during**: the number of questions asked by male attendees *during* the seminar (i.e. interrupting questions).

***F_during**: the number of questions asked by female attendees *during* the seminar (i.e. interrupting questions).

***question_time_start**: the time at which the questions start as HHMM e.g. 1650

***question_time_end**: the time at which the questions end as HHMM e.g. 1703

These two questions are to assess the duration of the question session and (roughly) for how long the speaker spoke.

The following questions are to control for the possible higher visibility of and (unconscious) preference given to more senior faculty in departments. We are defining *senior faculty* as the equivalents of University Teaching Officers, Academic Staff, etc. To collect these data in a standardised way, please go to the seminar department's website and collect these data from the "People" webpage (most departments have one) where they list "Academics" / "Teaching staff" / equivalent level of low-turnover, highly visible (e.g. teaching) staff.

**M_faculty**: the number of male senior faculty employed in the department

**F_faculty**: the number of female senior faculty employed in the department

We've made a small data collection sheet for you for the data that are to be collected during talks in case it helps (attached, with a particularly leading example in the top box). If you use any abbreviations, please be consistent so that these terms can be analysed straightforwardly in R.

Thanks very much for your help!

Alecia Carter
Dieter Lukas
Gillian Sandstrom
Alyssa Croft

| | | | | | |
|---|---|---|---|---|---|
| Date: | 16/06/2016 | Speaker: | Dieter Lukas | | |
| Talk title: | Mammals are awesome. | | | | |
| N_attendees: | 33 | N_male: | 17 | N_female: | 16 |
| First_question: | M | M_question: | HHH / | F_question: | /// |
| question_start_time | 1720 | question_end_time | 1732 | M_during: | / |
| | | | | F_during: | - |

---

| | | | | | |
|---|---|---|---|---|---|
| Date: | | Speaker: | | | |
| Talk title: | | | | | |
| N_attendees: | | N_male: | | N_female: | |
| First_question: | | M_question: | | F_question: | |
| question_start_time | | question_end_time | | M_during: | |
| | | | | F_during: | |

---

| | | | | | |
|---|---|---|---|---|---|
| Date: | | Speaker: | | | |
| Talk title: | | | | | |
| N_attendees: | | N_male: | | N_female: | |
| First_question: | | M_question: | | F_question: | |
| question_start_time | | question_end_time | | M_during: | |
| | | | | F_during: | |

---

| | | | | | |
|---|---|---|---|---|---|
| Date: | | Speaker: | | | |
| Talk title: | | | | | |
| N_attendees: | | N_male: | | N_female: | |
| First_question: | | M_question: | | F_question: | |
| question_start_time | | question_end_time | | M_during: | |
| | | | | F_during: | |

# Women's visibility in academic seminars: women ask fewer questions than men

Alecia Carter, Alyssa Croft, Dieter Lukas, Gillian Sandstrom

**Supplementary Table 1:**
Responses of a sample of academics who identify as male and female about what factors prevent them from asking a question after a seminar, even when they had a question to ask

| | | Reported importance | | | | | Kruskal-Wallis test | |
|---|---|---|---|---|---|---|---|---|
| Reason<br>Gender | N | Not at all (N, prop) | Slightly (N, prop) | Moderately (N, prop) | Very (N, prop) | Extremely (N, prop) | $\chi^2$ | *P* |
| Couldn't work up the nerve | | | | | | | 16.63 | <0.001 |
| Female | 277 | 39 (0.14) | 56 (0.20) | 52 (0.19) | 77 (0.28) | 53 (0.19) | | |
| Male | 188 | 54 (0.29) | 36 (0.19) | 40 (0.21) | 38 (0.2) | 20 (0.11) | | |
| The speaker was too eminent/intimidating | | | | | | | 17.15 | <0.001 |
| Female | 275 | 83 (0.30) | 92 (0.33) | 60 (0.22) | 31 (0.11) | 9 (0.03) | | |
| Male | 188 | 88 (0.47) | 59 (0.31) | 29 (0.15) | 9 (0.05) | 3 (0.02) | | |
| Not my field | | | | | | | 4.78 | 0.03 |
| Female | 276 | 44 (0.16) | 79 (0.29) | 70 (0.25) | 67 (0.24) | 16 (0.06) | | |
| Male | 186 | 46 (0.25) | 54 (0.29) | 43 (0.23) | 28 (0.15) | 15 (0.08) | | |
| Worried that I was not clever enough to ask a good question | | | | | | | 21.34 | <0.001 |
| Female | 276 | 55 (0.20) | 58 (0.21) | 62 (0.22) | 52 (0.19) | 49 (0.18) | | |
| Male | 188 | 75 (0.40) | 36 (0.19) | 33 (0.18) | 27 (0.14) | 17 (0.09) | | |
| Worried that I had misunderstood the question | | | | | | | 17.29 | <0.001 |
| Female | 276 | 18 (0.07) | 50 (0.18) | 60 (0.22) | 96 (0.35) | 52 (0.19) | | |
| Male | 187 | 19 (0.10) | 59 (0.32) | 36 (0.19) | 59 (0.32) | 14 (0.07) | | |
| Not sure whether the question was appropriate | | | | | | | 12.87 | <0.001 |
| Female | 276 | 21 (0.08) | 52 (0.19) | 89 (0.32) | 80 (0.29) | 34 (0.12) | | |
| Male | 187 | 27 (0.14) | 49 (0.26) | 54 (0.29) | 49 (0.26) | 8 (0.04) | | |
| I was meeting the speaker later / asked after the talk had ended | | | | | | | 2.02 | 0.16 |
| Female | 273 | 73 (0.27) | 71 (0.26) | 63 (0.23) | 51 (0.19) | 15 (0.05) | | |
| Male | 187 | 46 (0.25) | 33 (0.18) | 58 (0.31) | 38 (0.20) | 12 (0.06) | | |
| Not enough time | | | | | | | 3.96 | 0.05 |
| Female | 273 | 38 (0.14) | 78 (0.29) | 74 (0.27) | 67 (0.25) | 16 (0.06) | | |
| Male | 188 | 24 (0.13) | 43 (0.23) | 48 (0.26) | 50 (0.27) | 23 (0.12) | | |

Presented are the questions; the numbers of respondents of each gender who answered the question (N); the numbers and proportions of each gender who responded that the indicated factor was not at all, slightly, moderately, very, and extremely important (N, prop) for preventing them from asking questions; and the results of a Kruskal-Wallis test (in all cases, df = 1) indicating whether there was a difference between the genders' responses, including the test statistic ($\chi^2$) and significance (*p*).

**Supplementary Table 2:**
Responses of a sample of academics who identify as male and female, and who indicated that they believed that men asked more questions than women, about what they believed prevented women from asking a question if they had one.

| | | Reported importance | | | | | Kruskal-Wallis test | |
|---|---|---|---|---|---|---|---|---|
| **Reason** **Gender** | ***N*** | **Not at all (N, %)** | **Slightly (N, %)** | **Moderately (N, %)** | **Very (N, %)** | **Extremely (N, %)** | **$\chi^2$** | ***P*** |
| Can't work up the nerve | | | | | | | 12.39 | <0.001 |
| Female | 178 | 8 (0.04) | 22 (0.12) | 55 (0.31) | 72 (0.4) | 21 (0.12) | | |
| Male | 85 | 8 (0.09) | 17 (0.2) | 35 (0.41) | 20 (0.24) | 5 (0.06) | | |
| Feel intimidated by the speaker | | | | | | | 1.98 | 0.16 |
| Female | 178 | 14 (0.08) | 41 (0.23) | 65 (0.37) | 45 (0.25) | 13 (0.07) | | |
| Male | 83 | 9 (0.11) | 24 (0.29) | 27 (0.33) | 20 (0.24) | 3 (0.04) | | |
| Feel they are not an expert | | | | | | | 32.69 | <0.001 |
| Female | 178 | 7 (0.04) | 14 (0.08) | 44 (0.25) | 85 (0.48) | 28 (0.16) | | |
| Male | 85 | 17 (0.2) | 14 (0.16) | 28 (0.33) | 23 (0.27) | 3 (0.04) | | |
| Believe that they are not clever enough to ask a good question | | | | | | | 16.05 | <0.001 |
| Female | 178 | 14 (0.08) | 37 (0.21) | 44 (0.25) | 63 (0.35) | 20 (0.11) | | |
| Male | 84 | 21 (0.25) | 17 (0.2) | 26 (0.31) | 17 (0.2) | 3 (0.04) | | |
| Worried that they misunderstand the content | | | | | | | 43.93 | <0.001 |
| Female | 178 | 13 (0.07) | 35 (0.20) | 60 (0.34) | 56 (0.31) | 14 (0.08) | | |
| Male | 84 | 32 (0.38) | 19 (0.23) | 26 (0.31) | 6 (0.07) | 1 (0.01) | | |
| Are unsure that their questions are appropriate | | | | | | | 18.15 | <0.001 |
| Female | 178 | 7 (0.04) | 35 (0.2) | 53 (0.3) | 67 (0.38) | 16 (0.09) | | |
| Male | 85 | 19 (0.22) | 19 (0.22) | 25 (0.29) | 19 (0.22) | 3 (0.04) | | |
| Ask questions after the seminar is over | | | | | | | 21.24 | <0.001 |
| Female | 175 | 15 (0.09) | 46 (0.26) | 53 (0.3) | 42 (0.24) | 19 (0.11) | | |
| Male | 80 | 23 (0.29) | 24 (0.3) | 22 (0.28) | 9 (0.11) | 2 (0.03) | | |

Presented are the questions; the numbers of respondents of each gender who answered the question (N); the numbers and proportions of each gender who responded that the indicated factor was not at all, slightly, moderately, very, and extremely important (N, %) for preventing women from asking questions; and the results of a Kruskal-Wallis test (in all cases, df = 1) indicating whether there was a difference between the genders' responses, including the test statistic ($\chi^2$) and significance (*p*).

**Supplementary Table 3:**
Responses of a sample of academics who identify as male and female about what factors would encourage them to ask more questions after a seminar

| | | | Reported importance | | | | | Kruskal-Wallis test | |
|---|---|---|---|---|---|---|---|---|---|
| **Reason** / **Gender** | ***N*** | ***M*** | **Not at all (N, %)** | **Slightly (N, %)** | **Moderately (N, %)** | **Very (N, %)** | **Extremely (N, %)** | **$\chi^2$** | ***P*** |
| Confidence | | | | | | | | 29.97 | <0.001 |
| Female | 277 | 3.81 | 18 (0.06) | 18 (0.06) | 56 (0.20) | 93 (0.34) | 92 (0.33) | | |
| Male | 188 | 3.12 | 32 (0.17) | 29 (0.15) | 47 (0.25) | 44 (0.23) | 36 (0.19) | | |
| A chance to ask in person | | | | | | | | 5.46 | 0.02 |
| Female | 276 | 3.57 | 10 (0.04) | 24 (0.09) | 78 (0.28) | 126 (0.46) | 38 (0.14) | | |
| Male | 186 | 3.35 | 10 (0.05) | 25 (0.13) | 59 (0.32) | 73 (0.39) | 19 (0.1) | | |
| Seniority | | | | | | | | 41.19 | <0.001 |
| Female | 275 | 3.48 | 23 (0.08) | 29 (0.11) | 71 (0.26) | 97 (0.35) | 55 (0.20) | | |
| Male | 187 | 2.71 | 41 (0.22) | 42 (0.22) | 47 (0.25) | 44 (0.24) | 13 (0.07) | | |
| A longer time to formulate question | | | | | | | | 3.45 | 0.06 |
| Female | 276 | 2.76 | 28 (0.1) | 74 (0.27) | 122 (0.44) | 41 (0.15) | 11 (0.04) | | |
| Male | 187 | 2.57 | 32 (0.17) | 49 (0.26) | 78 (0.42) | 23 (0.12) | 5 (0.03) | | |
| Moderator doing a better job engaging whole audience | | | | | | | | 5.65 | 0.02 |
| Female | 271 | 2.71 | 43 (0.16) | 73 (0.27) | 91 (0.34) | 48 (0.18) | 16 (0.06) | | |
| Male | 185 | 2.45 | 51 (0.28) | 44 (0.24) | 54 (0.29) | 28 (0.15) | 8 (0.04) | | |
| Nicer speakers | | | | | | | | 2.31 | 0.13 |
| Female | 273 | 2.48 | 52 (0.19) | 88 (0.32) | 90 (0.33) | 37 (0.14) | 6 (0.02) | | |
| Male | 188 | 2.35 | 47 (0.25) | 63 (0.34) | 50 (0.27) | 22 (0.12) | 6 (0.03) | | |
| More welcoming host | | | | | | | | 5.83 | 0.02 |
| Female | 274 | 2.44 | 64 (0.23) | 82 (0.3) | 80 (0.29) | 40 (0.15) | 8 (0.03) | | |
| Male | 185 | 2.20 | 54 (0.29) | 69 (0.37) | 37 (0.2) | 21 (0.11) | 4 (0.02) | | |
| Having a moderator to ask the questions | | | | | | | | 24.03 | <0.001 |
| Female | 276 | 2.49 | 60 (0.22) | 80 (0.29) | 92 (0.33) | 30 (0.11) | 14 (0.05) | | |
| Male | 188 | 2.01 | 85 (0.45) | 46 (0.24) | 35 (0.19) | 15 (0.08) | 7 (0.04) | | |

Presented are the questions; the numbers of respondents of each gender who answered the question (N); the mean (*M*) of the responses of each gender (ordered from highest to lowest mean, averaged across gender); the numbers and proportions of each gender who responded that the indicated factor was not at all, slightly, moderately, very, and extremely important (N, %) for encouraging them to ask more questions; and the results of a Kruskal-Wallis test (in all cases, df = 1) indicating whether there was a difference between the genders' responses, including the test statistic ($\chi^2$) and significance (*p*).

# Women's visibility in academic seminars: women ask fewer questions than men

Alecia Carter, Alyssa Croft, Dieter Lukas, Gillian Sandstrom

**Supplementary Material 4**

More senior researchers (faculty) reported asking questions after seminars at higher frequencies than more junior researchers (post-graduate students and postdoctoral researchers; because of the small sample of undergraduate ($N$ = 12), research fellow ($N$ = 26), and "other" ($N$ = 27) respondents, these individuals' responses have been excluded from these analyses) (linear model with frequency of asking questions as the response and career stage as a predictor (faculty as the reference category): post-graduate students β ± S.E. = -0.61 ± 0.10, $t$ = -5.95, $p$ <0.001; postdoctoral researchers β ± S.E. = -0.40 ± 0.12, $t$ = -3.28, $p$ = 0.001; Fig 1a). However, these data show that men self-report asking questions more frequently than women (LM with women as the reference category: β ± S.E. = 0.38 ± 0.09, $t$ = 4.06, $p$ <0.001; Fig 1b) and that this relationship holds at the same rate at every career stage (no significant interaction between gender and career stage; Fig 1c).

Fig 1: The relationship between seniority and gender on the self-reported frequencies of question asking behaviour of the survey respondents

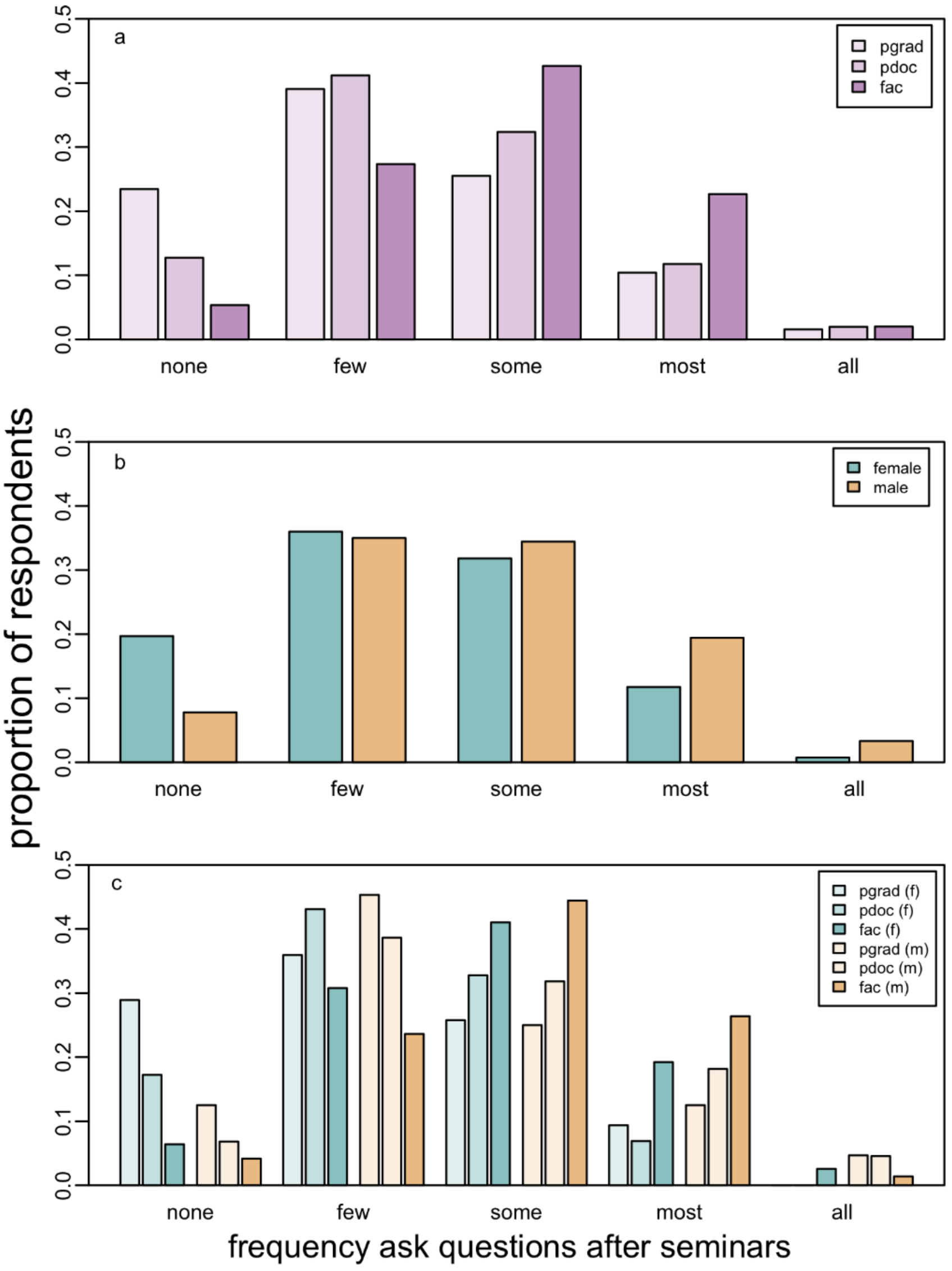

Fig 1. Shown are the proportions of respondents who report asking questions at seminars (a) at different career stages (pgrad = post-graduate student; pdoc = postdoctoral researcher; fac = faculty); (b) for respondents who identified as female and male; and (c) for female (f) and male (m) respondents at each career stage.